\documentclass[useAMS,usenatbib]{mn2e}

\usepackage{aas_macros}

\usepackage{graphicx}
\usepackage{amssymb}
\usepackage{amsmath}

\renewcommand{\vec}[1]{ {\bmath #1} } 

\newcommand{\hMsol}{{\>h^{-1}\rm M}_\odot}             
\newcommand{\hMpc}{{\>h^{-1}\rm  Mpc}}
\newcommand{\hkpc}{{\>h^{-1}\rm  kpc}}  
\newcommand{\kms}{\>{\rm   km\,s^{-1}}}  

\title[EGB radiation from dark matter annihilation]{Extragalactic gamma-ray background
  radiation from dark matter annihilation}
\author[J. Zavala, V.~Springel and M. Boylan-Kolchin]{\parbox{18cm}{Jes\'us
    Zavala$^{1,2}$\thanks{e-mail: jesus@mpa-garching.mpg.de}, Volker Springel$^{1}$
    and Michael Boylan-Kolchin$^{1}$\vspace{0.3cm}}\\ 
$^{1}$Max-Planck-Institut f\"{u}r Astrophysik, Karl-Schwarzschild-Stra\ss{}e 1, 85740 Garching
bei M\"{u}nchen, Germany\\
$^{2}$MPA/SHAO Joint Center for Astrophysical Cosmology at Shanghai
Astronomical Observatory, Nandan Road 80, Shanghai 200030, China}

\begin{document}



\maketitle

\label{firstpage}

\begin{abstract}
If dark matter is composed of neutralinos, one of the most exciting
prospects for its detection lies in observations of the gamma-ray
radiation created in pair annihilations between neutralinos, a process
that may contribute significantly to the extragalactic gamma-ray
background (EGB) radiation.  We here use the high-resolution
Millennium-II simulation of cosmic structure formation to produce the
first full-sky maps of the expected radiation coming from
extragalactic dark matter structures.  Our map making procedure takes
into account the total gamma-ray luminosity from all haloes and their
subhaloes, and includes corrections for unresolved components of the
emission as well as an extrapolation to the damping scale limit
of neutralinos.  Our analysis also includes a proper normalization of
the signal according to a specific supersymmetric model based on
minimal supergravity.  The new simulated maps allow a study of the
angular power spectrum of the gamma-ray background from dark matter
annihilation, which has distinctive features associated with the
nature of the annihilation process and may be detectable in
forthcoming observations by the recently launched FERMI satellite. Our
results are in broad agreement with analytic models for the gamma-ray
background, but they also include higher-order correlations not
readily accessible in analytic calculations and, in addition, provide
detailed spectral information for each pixel. In particular, we find
that difference maps at different energies can reveal cosmic
large-scale structure at low and intermediate redshifts. If the
intrinsic emission spectrum is characterized by an emission peak,
cosmological tomography with gamma ray annihilation radiation is in
principle possible.
\end{abstract}

\section{Introduction}

The nature of dark matter in the Universe is one of the most fascinating and
important mysteries of astrophysics today. Based on the evidence gathered
during the last decades, we know that a suitable particle physics candidate
has to be neutral and stable (or with a lifetime comparable to the age of the
Universe) to be the major component of dark matter, and that its interactions
with ordinary matter have to be very weak in order to be consistent with
constraints from cosmic structure formation.

Among the different theories that propose a solution to the dark
matter problem, supersymmetric (SUSY) extensions of the standard
particle physics model with R-parity conservation are particularly
attractive \citep[see for example][for a pedagogical introduction to
  SUSY]{Martin-98}. They generically predict the existence of a
lightest supersymmetric particle (LSP), which appears as a plausible
dark matter candidate. Among the possible LSPs resulting from SUSY
theories, two are strongly motivated and have been studied in detail:
the gravitino, which is the spin-3/2 superpartner of the graviton, and the
lightest neutralino, the spin-1/2 Majorana fermion that appears in the Minimal
Supersymmetric Standard Model (MSSM) as the lightest mass
eigenstate of a mixture of the superpartners of the neutral gauge
bosons (the fermions $\tilde{B}^0$ and $\tilde{W}^0$) and the neutral
SUSY Higgs ($\tilde{H}^0_u$ and $\tilde{H}^0_d$). Although gravitinos
are interesting viable candidates to be the major component of
dark matter (see for instance \citet{Feng-05,Steffen-06} for recent
reviews), they can be classified as {\em extremely} weakly interacting
particles \citep{Steffen-07} and are therefore very hard to detect
experimentally.  On the other hand, the character of neutralinos as
Majorana fermions (they are their own antiparticle) and their somewhat
stronger interactions as just weakly interacting massive particles (WIMPs)
make them more promising candidates for detection, possibly in the
near future.

Being WIMPs with expected masses of the order of $100\,{\rm GeV}$,
neutralinos have all the properties required to be the cold dark
matter (CDM) in the prevailing $\Lambda{\rm CDM}$ cosmogony, which is
at present the most successful model of structure formation
\citep[e.g.][]{Komatsu-09,Springel-Frenk-White-06}.  Significant
efforts are therefore undertaken to find experimental proof for the
existence of neutralinos. These experiments are usually classified as
direct and indirect searches \citep[for recent reviews on direct and
  indirect dark matter searches in the Galaxy
  see][]{Bertone-07,Hooper-07,Spooner-07, deBoer-08}.

Direct dark matter search experiments look for the recoil of ordinary
matter by scattering of WIMPs in laboratories on Earth.  Some of the
experiments that have been looking for signals are: CDMS
\citep{Akerib-04}, XENON \citep{Angle-08}, ZEPLIN \citep{Alner-05} and DAMA
\citep{Bernabei-00,Bernabei-08}.  The latter is the only experiment
that has reported a positive signal so far, a finding that 
has not yet been confirmed by another experiment, however.

The interaction rate of neutralinos with ordinary nuclei fundamentally
depends on two factors, the flux of neutralinos on Earth and the
interaction cross section of neutralinos with a given nuclei. The
latter can be computed theoretically within the context of the SUSY
model, but the former depends strongly on the phase space distribution
of dark matter at the Solar circle, which needs to be predicted by
cosmological models of structure formation.  In standard analysis, a
constant dark matter density and a simple Maxwell-Boltzmann
distribution for the particle velocities is often assumed, properly
modified to include the motion of the Sun around the Galaxy and that
of the Earth around the Sun \citep{Freese-Gondolo-Stodolsky-01}.
However, it is not at all clear that this simple assumption is
justified, because cold dark matter is expected to produce a large
number of overlapping cold streams locally, and may leave residual
structure in energy space. High-resolution N-body simulations of
structure formation combined with an explicit modelling of the local
stream density evolution provide one promising new approach to provide
a more accurate characterization of the possible signal
\citep{Vogelsberger-08,Vogelsberger-09b}.  Other dynamical local Solar
System effects may also be important \citep{Lundberg-Edsjo-04}.

Indirect dark matter searches look for the products of WIMP
annihilations generated in cosmic structures. Such possible
annihilation products include, for example, positrons
\citep{Baltz-Edsjo-99}, neutrinos \citep{Berezinsky-96} and gamma-ray
photons.  Interestingly, the satellite PAMELA has recently reported an
anomalously high positron abundance in cosmic rays at high energies
\citep{Adriani-09}.  Although this signal could be explained by a
number of astrophysical sources, dark matter annihilations offer a
particularly attractive explanation under certain scenarios
\citep{Bergstrom-Bringmann-Edsjo-08}, and this possibility has created
a flurry of activity in the field. In a related experiment, the FERMI
satellite has measured an excess in the total flux of electrons and positrons
in cosmic rays \citep{Abdo-09}, which, although is smaller than those
reported by the balloon experiments ATIC and PPB-BETS
\citep{Chang-08,Torii-08} is nevertheless significant.

Within our Galaxy, the annihilation signal is expected to be
particularly strong in the high density regions around the galactic
centre \citep{Bertone-Merritt-05a,Jacholkowska-06}. However, since
other astrophysical sources contribute prominently to the gamma-ray
flux there, it may be difficult to disentangle a possible neutralino
annihilation signal from the rest of
``ordinary'' sources.  A more promising approach may lie in an
analysis of the gamma-ray radiation produced in the smooth Milky Way
halo as a whole \citep{Stoehr-03,Fornengo-Pieri-Scopel-04,deBoer-04,Springel-08b}.
Particular attention has also been given to substructures within the
MW halo \citep[e.g.][]{Siegal-Gaskins-08,Ando-09}, which have been suggested to dominate the annihilation signal
observed at Earth \citep{Berezinsky-Dokuchaev-Eroshenko-03}, even
though they have recently been shown to be more difficult to detect
compared with the signal from the galactic centre
\citep{Springel-08b}.  Searches in other nearby galaxies have also
been considered, such as Andromeda \citep{Falvard-04,Lavalle-06} and
other galaxies in the Local Group \citep{Wood-08}. Clusters of
galaxies may also be interesting targets for searches
\citep{Jeltema-Kehayias-Profumo-08} due to their large content of dark
matter, even though a variety of other possible sources of high energy
gamma-rays in clusters, such as cosmic rays generated in structure
formation shocks, active galactic nuclei and supernovae, can lead to
confusion and may make it challenging to cleanly separate the dark
matter signal from the confusing sources.

On still larger scales, we expect that traces of dark matter
annihilation should be present in the extragalactic gamma-ray
background (EGB) radiation, which contains gamma-rays coming from all
possible cosmic sources.  In the present work, we will focus on these
{\em extragalactic} gamma-ray photons.  The EGB has been measured by
different gamma-ray satellites, in particular in the energy range
between $1\,{\rm MeV}$ and $30\,{\rm GeV}$ by COMPTEL and EGRET
\citep{Strong-Moskalenko-Reimer-04}. In the near future, our
understanding of the EGB will greatly improve by the Large Area
Telescope aboard the recently launched FERMI satellite \citep[formerly
  known as GLAST,][]{Atwood-09}, which covers an energy range between
$20\,{\rm MeV}$ and $300\,{\rm GeV}$ and features a greatly improved
sensitivity compared with its predecessor EGRET.  Several sources of
different astrophysical origin are known to contribute significantly
to the EGB, including blazars \citep{Ando-07b,Kneiske-Mannheim-08},
type Ia supernovae \citep{Strigari-05}, cosmic rays accelerated at
cosmological structure formation shocks
\citep{Miniati-02,Miniati-Koushiappas-DiMatteo-07,Jubelgas-08}, among
others \citep[see][for recent reviews]{Dermer-07,Kneiske-08}.  Of
course, dark matter annihilation also contributes to the EGB, and it
is in this component that we focus on in this work.

Using standard analytic approaches to describe cosmic structure
formation, like halo mass functions and clustering amplitude as a
function of redshift, it is possible to make predictions for the
relative gamma-ray flux coming from haloes and their subhaloes as a
function of mass and redshift \citep{Ullio-02,Taylor-Silk-03} and to use
this information to estimate the contribution of dark matter
annihilation to the EGB.  Both, the isotropic and anisotropic
components of this radiation have been predicted in this way in
previous works
\citep{Elsasser-Mannheim-05,Ando-Komatsu-06,Cuoco-07,Cuoco-08,deBoer-07,Fornasa-09}.
In particular, \citet{Ando-Komatsu-06} first pointed out that the
density squared dependence of dark matter annihilation should
leave a specific signature in the angular power spectrum of the EGB.
We also note that \citet{Siegal-Gaskins-Pavlidou-09} have shown that the
energy dependence of the angular power spectrum of the total diffuse gamma-ray
emission can be used to disentangle different source
populations, including that associated with dark matter annihilation.

However, it is not clear how accurately these analytic approaches
capture the non-linear structures resolved in the newest generation of
high-resolution cosmological simulations. Also, the previous analysis
have so far been restricted to statistical statements about the power
spectrum of the EGB, or its mean flux. For a full characterization of
the signal, it would be useful to have accurate realizations of maps
of the expected gamma-ray emission over the whole sky. Such maps
contain in principle the full information that can be harvested, and
allow, for example, correlation studies with foreground large-scale
structure in terms of galaxies or dark matter.

In the present paper, we therefore focus on techniques to predict the
extragalactic gamma-ray background directly from cosmological N-body
simulations. To this end, we use the Millennium-II simulation. With
$10^{10}$ particles in a homogeneously sampled volume of
$(100\,h^{-1}{\rm Mpc})^3$, it is one of the best resolved structure
formation simulations to date. We use a ray-tracing technique to
accumulate the signal from all haloes and subhaloes over the past
light-cone of a fiducial observer, positioned at a plausible location of
the Milky Way in the simulation box. Through careful accounting for
unresolved structure, we can extrapolate our signal to the damping scale 
limit and obtain the first realization of realistic full sky maps of
the expected gamma-ray radiation from dark matter annihilation. The
maps can be used both to check earlier analytic estimates, and to
develop new strategies for identification and exploitation of the
annihilation component in the EGB. For example, we show that
difference maps at different energies could be used to uncover nearby
cosmic large-scale structure, which could be correlated with other
probes of large-scale structure to more unambiguously identify the
origin of this background component.

This paper is organized as follows.  In Section 2, we briefly review
the analytic formulae associated with the contribution to the EGB from
dark matter annihilation.  In Section 3, we describe the SUSY model
chosen for our analysis.  We then present the N-body simulation used,
and the effect of dark matter clustering in the annihilation's gamma
ray flux in Section 4.  In Section 5 we describe our method to
generate simulated sky maps of the EGB radiation from dark matter
annihilation, and we analyse the energy and angular power spectra of
the signal. We give a summary and our conclusions in Section~6.

\section{Dark matter annihilation rates and the EGB}

If dark matter is made of particles that can annihilate with each
other, then the number of annihilations per unit time in a given
volume $V$ is given by \citep[e.g.][]{Gondolo-Gelmini-91}:

\begin{equation}\label{rate}
n=\frac{\langle\sigma v\rangle
}{2m^2_{\chi}}\int_V\rho^2_{\chi}(\vec{x})\,{\rm d}^3x,
\end{equation}
where $m_{\chi}$ and $\rho_{\chi}$ are the mass and density of the dark matter
species, and $\langle\sigma v\rangle$ is the thermally averaged product of the
annihilation cross section and the M\o ller velocity.

Since the particle physics details related to the intrinsic nature of
dark matter are encoded in the term outside the integral in
Eq.~(\ref{rate}), it is convenient to remove this dependence in
analyzing the effect of dark matter clustering on the annihilation
rate. For this purpose, it is customary to introduce the ratio of
gamma ray emission coming from a given halo of volume $V$ to the
average emission of a homogeneous distribution of the same amount of
dark matter within this volume:
\begin{equation}\label{flux}
f_h(V)=\frac{1}{\bar{\rho}_h^2V}\int_{V}\rho^2_{\chi}(\vec{x})\,{\rm
  d}^3x
\end{equation} 
where $\bar{\rho}_h$  is the average  density of dark matter  particles inside
$V$. 

For instance if the region of interest is a dark matter halo, then
Eq.~(\ref{flux}) can be solved analytically by assuming a specific
density distribution.  In particular, for the spherically symmetric NFW
profile \citep{Navarro-Frenk-White-97}, $f_h(V)$ is given by
\citep{Taylor-Silk-03}:
\begin{equation}\label{flux_nfw}
f_h(c_{\Delta})=\frac{c_{\Delta}^3\left[1-1/(1+c_{\Delta}^3)\right]}{9\left[\textrm{ln}(1+c_{\Delta})-c_{\Delta}/(1+c_{\Delta})\right]^2},
\end{equation}
where $c_{\Delta}=r_{\Delta}/r_s$ is the concentration parameter,
$r_s$ is the scale radius of the NFW profile, and $r_{\Delta}$ is the
limiting radius of the halo, connected to its enclosed mass $M_{\Delta}$
through:
\begin{equation}\label{mass_delta}
M_{\Delta}=\frac{4}{3}\pi\bar{\rho}_h
r_{\Delta}^3=\frac{4}{3}\pi\Delta\rho_{\rm crit} r_{\Delta}^3,
\end{equation}
where $\Delta$ is the overdensity parameter. Different values for
$\Delta$ are used in the literature, with the two main choices being
$\Delta=200$ \citep{Navarro-Frenk-White-97}, and
$\Delta\sim178\Omega_m^{0.45}$ for a $\Lambda$CDM model with
$\Omega_m+\Omega_\Lambda=1$ \citep{Eke-Navarro-Steinmetz-01}.

For spherically symmetric haloes, 
the integral in Eq.~(\ref{flux}) can also be written as the scaling law
 \citep{Springel-08b} $L_h' \propto
V_{\rm max}^4 / r_{\rm half}$, where $V_{\rm max}$ is the maximum rotational
velocity of the halo and $r_{\rm half}$ is the radius containing half of
the gamma-ray flux. For a NFW halo, the latter formula reduces to
\begin{equation}\label{scaling_law}
L_h'=\int\rho^2_{\rm NFW}(r)\,{\rm d}V=\frac{1.23\,V_{\rm
    max}^4}{G^2r_{\rm max}},
\end{equation}
where $r_{\rm max}$ is the radius where the rotation curve reaches its
maximum. We will use Eq.~(\ref{scaling_law}) for our analysis in this
paper.  The most recent high-resolution N-body simulations support a
slightly revised density profile that becomes gradually shallower
towards the centre \citep{Navarro-08}, deviating from the asymptotic
power-law cusp of the NFW profile. For this Einasto profile, the
coefficient $1.23$ in Eq.~(\ref{scaling_law}) changes to 1.87, an
insignificant difference compared to other uncertainties in the
analysis of the dark matter annihilation radiation.

We can extend Eq.~(\ref{flux}) to larger volumes containing several
haloes of different sizes. If we assume that all dark matter particles
in a given cosmic volume $V_B$ belong to virialized haloes, then the
ratio of gamma-ray emission coming from all haloes contained in $V_B$
with masses larger than $M_{\rm min}$ to the emission produced by a
smooth homogeneous distribution of dark matter, with average density
$\bar{\rho}_B$, filling this volume is given by:
\begin{equation}\label{flux_mmin}
f(M>M_{\rm min})=\frac{1}{\bar{\rho}_B^2V_B}\int_{M_{\rm
    min}}^{\infty}{\rm d}N_{V_B}(M)\bar{\rho}_h^2V(M)f_h(V(M))
\end{equation} 
where ${\rm d}N_{V_B}(M)$ is the number of  haloes in the volume $V_B$ with masses in
the range $[M,M+{\rm d}M]$, which can be written as
\begin{equation}\label{n_halos}
{\rm
  d}N_{V_B}(M)=V_B\frac{\textrm{d}n(M)}{\textrm{d}M}\textrm{d}M=V_B\frac{\textrm{d}n(M)}{\textrm{dlog}M}\textrm{dlog}M ,
\end{equation}
with   $\textrm{d}n(M)/\textrm{dlog}M$   denoting   the  differential   halo   mass
function. Thus:
\begin{eqnarray}\label{flux_mmin_2}
f(M>M_{\rm min})=\ \ \ \ \ \ \ \ \ \ \ \ \ \ \ \ \ \ \ \ \ \ \ \ \ \ \ \ \ \ \ \ \ \ \ \ \ \ \ \ \ \ \ \ \  \nonumber\\
\frac{1}{\bar{\rho}_B^2}\int_{M_{\rm
    min}}^{\infty}\left(\frac{\textrm{d}n(M)}{\textrm{dlog}M}\right)\bar{\rho}_h^2V(M)f_h(V(M))\textrm{dlog}M .
\end{eqnarray} 

In the literature, $f(M>M_{\rm min})$ has been called the ``flux
multiplier'' \citep[Eq.~18 of][]{Taylor-Silk-03} or the ``clumping
factor'' \citep[Eq.~6 of][]{Ando-Komatsu-06}.  The limiting lower mass
in Eq.~(\ref{flux_mmin_2}) is the minimum mass scale for virialized
haloes surviving today, which is roughly given by the free streaming
length of dark matter particles: $\sim10^{-6}M_{\odot}$ for
$m_\chi=100\,{\rm GeV}$
\citep{Hofmann-Schwarz-Stocker-01,Green-Hofmann-Schwarz-04},
see section 4.1 for a discussion on this cutoff mass.  Although many
of the small scale haloes have suffered from tidal truncation or were
disrupted, it is possible that they contribute significantly to the
gamma-ray emission produced by dark matter annihilation.

Eqs. (\ref{flux}) and (\ref{flux_mmin_2}) allow us to study the gamma ray flux
coming  from individual haloes  or groups  of haloes  in a  given region  of the
Universe at  a given redshift (both  can be applied to  any redshift, provided
that  $\rho_{\chi}$, $\bar{\rho}_h$  and $\bar{\rho}_B$  are computed  at such
redshift).

Following Eq.~(\ref{rate}), we define the gamma ray emissivity (energy
emitted per unit energy range, unit volume and unit time) associated
with dark matter annihilation as
\begin{equation}\label{emiss}
\epsilon_{\gamma}=E_{\gamma}\frac{{\rm d}N_{\gamma}}{{\rm d}E_{\gamma}}\frac{\langle\sigma
v\rangle}{2}\left[\frac{\rho_{\chi}}{m_{\chi}}\right]^2,
\end{equation}
where  ${\rm d}N_{\gamma}/{\rm d}E_{\gamma}$  is the  total  differential photon  spectrum
summed  over all  channels  of annihilation.  Of  all the  quantities in  this
equation,  only $\rho_{\chi}$  depends  on the  spatial  distribution of  dark
matter, the rest is related to its intrinsic properties. For the analysis of
the latter, since we consider
dark matter to be made of neutralinos within a SUSY scenario, we
conveniently   define   the   so  called   SUSY
factor\footnote{This  term  has  been   extensively  used  previously  in  the
  literature, for  example in \citet{Fornengo-Pieri-Scopel-04},
  $f_{\rm SUSY}$ is
  related to the integral quantity $\Phi^{\rm SUSY}$ defined there.}:
\begin{equation} \label{fsusy}
f_{\rm SUSY}= \frac{{\rm d}N_{\gamma}}{{\rm
    d}E_{\gamma}}\frac{\langle\sigma v\rangle}{m^2_{\chi}} .
\end{equation}

The signal that we are ultimately interested in is the contribution of
neutralino annihilation to the EGB, or more specifically, the spatial
distribution of the gamma-ray radiation coming from dark matter
annihilation integrated over all redshifts along the line-of-sight for
a fiducial observer, located at $z=0$, for all directions on its
two-dimensional full sky. The values we want to calculate for each
pixel of such a sky map are that of the specific intensity, the
energy of photons received per unit area, time, solid angle and energy
range:
\begin{equation}\label{intensity}
I_{\gamma,0}=\frac{1}{4\pi}\int
\epsilon_{\gamma,0}(E_{\gamma,0}(1+z),z)\frac{{\rm d}r}{(1+z)^4},
\end{equation}
where the integral is over the whole line of sight, $r$ is the
comoving distance and $E_{\gamma,0}$ is the energy measured by the
observer at $z=0$. Note that $\epsilon_{\gamma,0}$ is evaluated at the
blueshifted energy $(1+z)E_{\gamma,0}$ along the line-of-sight to
compensate for the cosmological redshifting.

\section{Dark matter annihilation and the SUSY factor}

In the present section we describe the particular SUSY model that we
used to compute the gamma-ray emissivity produced by dark matter
annihilation. In particular, we provide a closer examination of the
SUSY factor given in Eq.~(\ref{fsusy}).

A detailed particle physics model for the nature of the neutralino as
dark matter, and its mass and interactions, is needed to compute
$f_{\rm SUSY}$.  We restrict our analysis to the framework of the
minimal supergravity model \citep[mSUGRA, for reviews
  see][]{Nilles-84, Nath-03}.  The mSUGRA model is popular, among
other reasons, due to its relative simplicity, which reduces the large
number of parameters in the general MSSM ($\gtrsim 100$) to
effectively four free parameters, $m_{1/2}$, $m_0$, tan$\beta$,
$A_{0}$ and the choice of sign for $\mu$.  The parameters $m_{1/2}$,
$m_0$ and $A_0$ are the values of the gaugino and scalar masses and
the trilinear coupling, all specified at the GUT scale, tan$\beta$ is
the ratio of the expectation values in the vacuum of the two neutral
SUSY Higgs ($\tilde{H}^0_u$ y $\tilde{H}^0_d$), and finally, $\mu$ is
the Higgsino mass parameter.

The general 5-dimensional parameter space of the mSUGRA model is
significantly constrained by various requirements: consistency with
radiative electroweak symmetry breaking (EWSB) and experimental
constraints on the low energy region, which are given by current lower
mass bounds for the lightest SUSY Higgs after EWSB, $m_h$, as well as
concordance between experimental and theoretical predictions for the
decay $b\rightarrow s\gamma$ and the anomalous magnetic moment of the
muon $(g_{\mu}-2)/2$. Importantly, it is normally also assumed that
the major component of dark matter is the LSP of the theory, then the
LSP must be neutral and $\Omega_{\rm LSP}=\Omega_{\rm DM}$, which reduces the
allowed regions in the parameter space considerably.  For typical
mSUGRA scenarios, the LSP that satisfies these conditions is the
lightest neutralino.  Its relic density $\Omega_\chi$ needs to be
properly calculated by solving the Boltzmann equation, with the
parameters of the theory chosen such that the desired dark matter
density results.

Observational constraints on the abundance of dark matter,
$\Omega_{\rm DM}$, come from various sources, including the cosmic
microwave background (CMB), galaxy clusters, supernovae,
Lyman-$\alpha$ forest data, and weak gravitational lensing.  For the
purposes of the analysis in this section, we take the $1\sigma$ bound
for the amount of dark matter in the Universe according to the five
year results of WMAP data combined with distance measurements from
type Ia supernovae and Baryon Acoustic Oscillations
\citep{Komatsu-09}:
\begin{equation} \label{WMAP5}
0.1109\leq\Omega_{\rm DM}h^2\leq0.1177.
\end{equation}

\begin{figure}
\centering
\includegraphics[height=8.0cm,width=10.5cm]{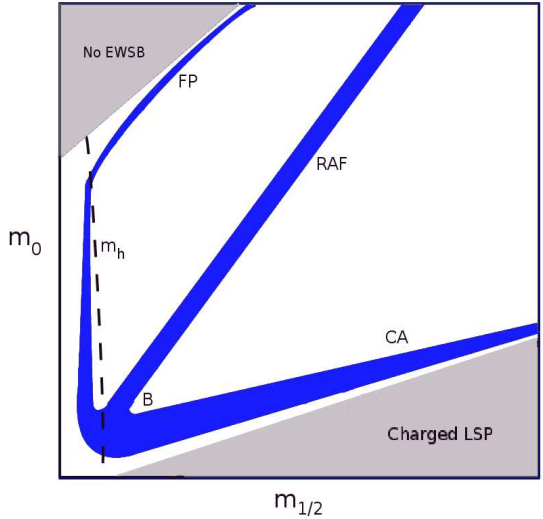}
\caption{Sketch showing the allowed parameter space of the mSUGRA
  model when the constraint on the relic density according to
  Eq.~(\ref{WMAP5}) (dark blue region) is imposed.  The experimental
  constraint on the mass of the lightest SUSY Higgs, $m_h$, is also
  shown for reference.  The grey region in the lower right corner is
  where the LSP is charged, whereas in the grey region in the upper
  left corner there is no radiative EWSB. The abbreviated names for
  various important regions in the parameter space are marked in the
  figure with capital letters, see text for details.}
\label{msugra}
\end{figure}

The general picture for the allowed regions in the parameter space of
the mSUGRA model resulting from the different constraints can be
qualitatively described in the 2-dimensional plane ($m_0-m_{1/2}$), as
shown in Fig.~\ref{msugra}.  The blue area in the figure represents
the allowed region according to Eq.~(\ref{WMAP5}) for the
observational bounds on the abundance of dark matter.

\begin{table*}
 \begin{minipage}{170mm}
   \centering
  \begin{tabular}{@{}lrrrrrrrrrrrr@{}}
  \hline
   Model\footnote{The letters to represent each point were chosen to
 follow the notation of \citet{Battaglia-04} and \citet{Gondolo-04}} & A &
 C & D & E & F & G & H & I & J & K & L \\
 \hline
 $m_{1/2}$  & 550  & 435 & 520 & 399 & 811 & 396 & 930
 & 395 & 750  & 1300  & 450 \\
 $m_{0}$  & 114 & 94 & 113 & 2977 & 4307 & 118.5 & 242.5
 & 193 & 299.5  & 1195  & 299 \\
 tan$\beta$  & 7 & 11 & 10 & 30 & 30.4 & 20 & 20
 & 35 & 35  & 46  & 47 \\
 sign($\mu$)  & $+$ & $+$ & $-$ & $+$ & $+$ & $+$ & $+$
 & $+$ & $+$  & $-$  & $+$ \\
 $m_{\chi}$  & 226.4 & 175.1 & 214.9 & 154.2 & 337 & 160.4 & 394.2
 & 160.6 & 315.5  & 566.1  & 184.9 \\
 $\Omega_{\chi}h^2$  & 0.115 & 0.114  & 0.116 & 0.118 & 0.112 & 0.113 & 0.118
 & 0.118 & 0.114  & 0.116  & 0.111 \\
 Region\footnote{B for Bulk region, FP for Focus Point region,
 CA for Co-annihilation region and RAF for Rapid Annihilation
 Funnel region}  & CA & B & CA & FP & FP & B & CA
 & B & CA  & RAF  & B \\
\hline
\end{tabular}
  \caption{Selected benchmark points in the mSUGRA parameter
    space. For our main analysis, we will focus on point `L', which
    predicts an annihilation flux close to the maximum possible for
    the constrained mSUGRA models. }
\end{minipage}
\end{table*}

The different regions in the mSUGRA parameter space that satisfy these
constraints have been studied abundantly in the past
\citep[e.g.][]{Baer-Balazs-03,Battaglia-04,Feng-05,Belanger-Kraml-Pukhov-05}
and have received generic names, marked with capital letters in the
figure: (B) bulk region (low values of $m_0$ and $m_{1/2}$), (FP)
focus point region (large values of $m_0$), (CA) co-annihilation region
(low $m_0$ and $m_{\chi}\lesssim m_{\tilde{\tau_1}}$, where
$\tilde{\tau_1}$ is the lightest slepton) and (RAF) rapid annihilation
funnel region (for large values of tan$\beta$ and a specific condition
for $m_{\chi}$\footnote{The relation which should hold is:
  $2m_{\chi}\approx m_{A}$, where $A$ is the CP-odd Higgs boson; see
  for example \citet{Feng-05} for details.}).  The regions in grey
(light) represent zones where the LSP is charged (lower right corner)
and where there is no radiative EWSB solution (upper left corner). The
dashed line is shown as a reference for the experimental constraints
for $m_h$.

Our goal in this section is to analyze the supersymmetric factor
$f_{\rm SUSY}$ predicted by the mSUGRA model in the different allowed
regions depicted in Fig.~\ref{msugra}.  To do so, we define several
``benchmarks points'' which fulfill all the experimental constraints
discussed above and which can be taken as representative of the
different allowed regions in parameter space that we just described.  The
benchmark points were selected following the analysis of
\citet{Battaglia-04} and \citet{Gondolo-04}, but slightly modified to
be consistent with the cosmological constraint of Eq.~(\ref{WMAP5}).
We use the numerical code {\small DarkSUSY}
\citep{Gondolo-04,Gondolo-05} with the interface {\small ISAJET}
\citep{Baer-03} to analyse these benchmark points.  Table~1 lists the
11 selected benchmark points with their corresponding values for the
five mSUGRA parameters, the value of the lightest neutralino's mass,
its relic density, and the name of the region they belong to in
Fig.~\ref{msugra}.

\begin{figure}
\centering
\includegraphics[height=8.5cm,width=8.5cm]{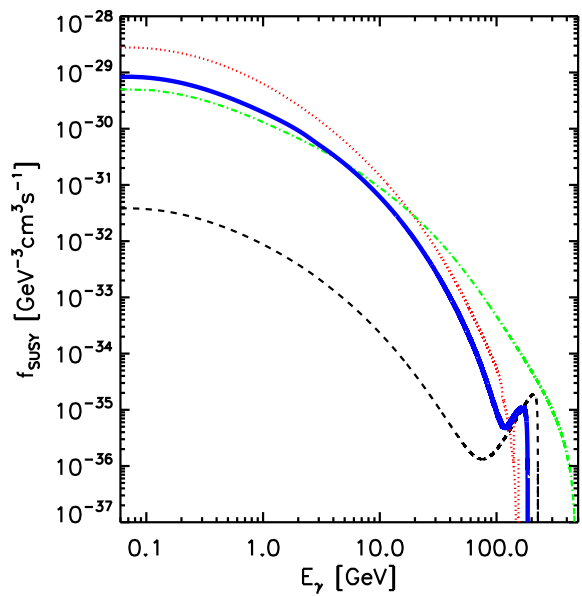}
\caption{Supersymmetric factor as a function of $\gamma$-ray emission
  energy for a selection of the benchmark points described in Table 1:
  A, E, K and L, which are shown with black (dashed), red (dotted), green
  (dash-dotted) and blue (solid) lines, respectively.}
\label{susy_energy}
\end{figure}

Fig.~\ref{susy_energy} shows the energy spectrum of $f_{\rm SUSY}$
computed for four of the benchmark points presented in Table~1: A
(black dashed line), E (red dotted line), K (green dash-dotted line), and L
(blue solid line), which are representatives
of the regions CA, FP, RAF and B, respectively. The main features of
Fig.~\ref{susy_energy} are determined by the photon spectrum ${\rm
  d}N_{\gamma}/{\rm d}E_{\gamma}$ which receives contributions from
three mechanisms of photon production \citep{Bringmann-Bergstrom-Edsjo-08}: 
(i) gamma-ray continuum emission following the decay of neutral pions produced during the
hadronization of the primary annihilation products at the tree-level;
this mechanism is dominant at low and intermediate energies (in
Fig.~\ref{susy_energy} it dominates the spectrum in most of the energy
range); (ii) monoenergetic gamma-ray lines for neutralino annihilation
in two-body final states containing photons, which is allowed in
higher order perturbation theory; the most relevant final states
containing these lines are $\chi\chi\rightarrow\gamma\gamma$ and
$\chi\chi\rightarrow Z^0\gamma$, where $Z^0$ is the neutral $Z$ boson;
for non-relativistic neutralinos the energy of each outgoing photons
in these processes is $E_{\gamma}=m_{\chi}$ and
$E_{\gamma}=m_{\chi}(1-m_{Z^0}^2/4m_{\chi}^2)$, respectively (in
Fig.~\ref{susy_energy} these lines are prominent for benchmark point E
(red line) and negligible in the rest of the cases); (iii) when
the final products of annihilation are charged, internal
bremsstrahlung (IB) needs to be considered, it leads to the emission
of an additional photon in the final state; the IB contribution to the
spectrum is typically dominant at high energies resulting in a
characteristic bump (benchmark points A and L have an important
contribution from IB at energies close to $m_{\chi}$, for the other
two points the contribution from IB is negligible). Processes (ii) and
(iii) are subdominant relative to process (i), but they display
distinctive spectral features intrinsic to the phenomenon of
annihilation. 

Additionally, the annihilation of neutralinos into electrons
and positrons contributes indirectly to the high energy photon spectrum when
low energy background photons, such as those in the cosmic microwave
background (CMB), are scattered to higher energies via inverse
Compton (IC) scattering by these electrons and positrons
\citep{Profumo-Jeltema-09,Belikov-Hooper-09}. In the case of CMB photons, the 
energy peak of this IC photon emission ($E_{\gamma}^{IC}$) is approximately
independent of redshift
and is given by: $E_{\gamma}^{IC}\approx
E_{\gamma}^{CMB}(E_{e^{\pm}}/m_{e^{\pm}}c^2)^2$, where $E_{\gamma}^{CMB}$
and $E_{e^{\pm}}$ are the average energies of the CMB photons and $e^{\pm}$
respectively. For neutralino masses in the $100~$GeV range, this energy
peak lies in the X-ray regime ($\sim10~$keV) \citep{Profumo-Jeltema-09}. Unless the
neutralino mass is larger than $1~$TeV and with a hard $e^{\pm}$
spectrum, the contribution of this mechanism is not significant in the
energy range relevant to the present study ($\gtrsim0.1~$GeV). We 
note that we are currently working on
the analysis of the X-ray extragalactic background radiation from dark matter
annihilation and we will present our results in a future work.
  
In Fig.~\ref{susy_energy}, we have shown only 4 of the 11 benchmark
points we considered; the rest show very similar features and lie in
between these 4 cases. Fig.~\ref{susy_energy} also indicates that the
normalization of the spectrum of $f_{\rm SUSY}$ typically varies by
three orders of magnitude among the different regions of the allowed
parameter space. Also there appears to be a trend between the
different allowed regions in the mSUGRA model: $f_{\rm SUSY}$ has the
largest value for benchmark points on the focus point region (FP, red
dotted line), followed by the bulk region (B, blue solid line), the rapid
annihilation funnel region (RAF, green dash-dotted line), and finally the
co-annihilation region (CA, black dashed line). This last feature could be of
importance for constraining the allowed mSUGRA parameter space even
more: in principle, a precise measurement of the gamma-ray flux coming
from dark matter annihilation could discriminate between the
characteristic regions FP, B, CA and RAF, considering for example that
the difference in $f_{\rm SUSY}$ between the CA and FP regions is
around two orders of magnitude. However, such inferences will only
become possible if the remaining sources of uncertainty can be reduced
to less than two orders of magnitude. This is a demanding goal, given
the limited knowledge we have on some of the other astrophysical
factors that play an important role in the production of these
gamma-rays, as well as the observational difficulties in properly
subtracting from an observed gamma-ray signal the contributions from
astrophysical sources unrelated to dark matter
annihilation. Nevertheless, `dark matter astronomy' remains an
interesting possibility in the light of the results shown in
Fig.~\ref{susy_energy}.

In our subsequent analysis, we will now choose one particular
benchmark point and adopt its $f_{\rm SUSY}$ value for the rest of our
work, keeping in mind the results above. Our chosen benchmark point is
the model L (that we highlight with a thick blue solid line in
Fig.~\ref{susy_energy}), which gives an upper limit on $f_{\rm SUSY}$
for the benchmark points in the bulk region and is close to the
maximum value of $f_{\rm SUSY}$ for all the benchmark points analyzed.
The latter is the main reason to choose this particular model because
it is desirable that $f_{SUSY}$, and hence the predicted gamma-ray flux,
has a large value among the different theoretical possibilities. This
optimistic choice as far as prospects for future detections are
concerned then also demarcates the boundary where non-detections can
start to constrain the mSUGRA parameter space. But there are other
reasons as well. The mass of the neutralino in this model is
``safely'' larger than the lower mass bounds coming from experimental
constraints \citep[$m_\chi>50\,{\rm GeV}$ according to][]{Heister-04},
but low enough to be detectable in the experiments available in the
near future. Also, a high value of the parameter tan$\beta$ seems to
be favored by other theoretical expectations \citep[e.g.][]{Nunez-08}.
We note that although benchmark point L does not have prominent
gamma-ray lines in its spectrum, it does have an important IB
contribution and serves the point of showing the implications of a
peak in the annihilation energy spectrum for the EGB.

We note that recent analyses on the cosmic-ray anomalies as measured by
PAMELA and FERMI favour higher neutralino masses ($\sim1$~TeV) 
when these anomalies are explained mainly by dark matter annihilation
\citep[e.g.][]{Bergstrom-Edsjo-Zaharijas-09}. However, it
is important to keep in mind that other astrophysical sources, such as
pulsars and supernovae remnants, are expected
to have a meaningful contribution to the solution of the cosmic ray
anomalies. At present, these experiments are only suggestive of dark matter
particles  in the TeV range, not definite. Our goal is to give an account of
the general features of the gamma-ray spectrum from annihilation and their
imprint on the EGB. For this purpose, our fiducial model with $m_{\chi}\sim200$~GeV is
adequate. Models with higher masses would shift the
energy scale to higher values, but our general conclusions would remain
unaltered.

\section{Gamma-ray luminosity from annihilation in haloes and subhaloes: The
  astrophysical factor}

In this section we analyse the astrophysical part of the annihilation
gamma-ray flux, arising from the clustering of dark matter into haloes and
subhaloes across cosmic time.  To this end we analyse the dark matter haloes
in one of the most recent state-of-the-art cosmological N-body simulations.

Specifically, we use the ``Millennium-II'' simulation (MS-II) of
\citet{Boylan-Kolchin-09} which has the same particle number ($2160^3$) and
cosmological parameters ($\Omega_m=0.25$, $\Omega_{\Lambda}=0.75$, $h=0.73$,
$\sigma_8=0.9$ and $n_{s}=1$, where $\Omega_m$ and $\Omega_{\Lambda}$ are the
contribution from matter and cosmological constant to the mass/energy density
of the Universe, respectively, $h$ is the dimensionless Hubble constant
parameter at redshift zero, $n_s$ is the spectral index of the primordial
power spectrum, and $\sigma_8$ is the rms amplitude of linear mass fluctuations
in $8\hMpc$ spheres at redshift zero) as the Millennium simulation (MS-I)
\citep{Springel-05b} but with a box size that is 5 times smaller, equal to
$L=100\hMpc$ on a side, thus having a mass resolution of
$6.89\times10^6\hMsol$, 125 times smaller than in MS-I.  Typical Milky-Way
sized haloes are resolved with several $10^5$ particles, while clusters of
galaxies have about 50 million particles.

For the dynamical evolution of the MS-II, a fixed comoving gravitational
softening length of $\epsilon=1\hkpc$ (Plummer-equivalent) was used. The MS-II
was processed on-the-fly to find haloes using a friend-of-friends (FOF)
algorithm, followed by a post-processing step to identify gravitationally
bound dark matter substructures down to a limit of $20$ particles within each
group based on the {\small SUBFIND} algorithm \citep{Springel-01}. A given FOF
halo is decomposed by {\small SUBFIND} into disjoint subgroups of self-bound
particles. The most massive of these subgroups represents the ``main'' or
``background'' halo of a given FOF halo \citep[see Fig.~3
of][]{Springel-01}. For the reminder of this work we will refer to these
subgroups as main haloes or simply haloes, and to the rest of the identified
subgroups of a given FOF halo as subhaloes of that main halo. {\small SUBFIND}
also computes several properties for each subgroup; the ones we will use
extensively in this work are $V_{\rm max}$ and $r_{\rm max}$, the maximum
circular velocity of a (sub)halo and the radius where this maximum is
attained. We also use $r_{1/2}$, the radius which encloses half of the mass of
a subhalo, and $r_{200}$, the radius where the mean density of a main halo is
200 times the mean background matter density.

Even with the large particle number available in the MS-II simulation,
it is important to note that a direct evaluation of halo luminosities
by estimating local dark matter densities at the positions of each
simulation particle (for example based on the SPH kernel interpolation
technique) would be seriously affected by resolution limitations.
This is because most of the emission originates over a tiny radial
region, close to the very centre of a halo, where the density is
easily underestimated even for otherwise well resolved structures.  In
fact, it requires of order $10^8 - 10^9$ particles to obtain a
converged estimate of the emission from the smooth part of a cuspy
dark matter halo in a brute-force approach \citep{Springel-08b}. On
the other hand, it is numerically {\em very much easier} to resolve
the existence of a halo or subhalo, and to accurately estimate its
total mass, as well as its primary structural properties such as
$V_{\rm max}$ and $r_{\rm max}$.  Our strategy will therefore be to
adopt an analytic fit for the density profile of each detected dark
matter halo and to predict its expected emission by integrating this
profile. In other words, we will use the scaling relation
(\ref{scaling_law}) to compute $L_h'$ for main haloes and subhaloes
alike.  Since $V_{\rm max}$ and $r_{\rm max}$ can be measured
reasonably accurately even for poorly resolved haloes, this gives us a
reliable accounting for the total emission of all haloes down to a
fairly low mass limit. Importantly, we can completely avoid in this
way to introduce strong mass-dependent resolution effects in our
measurements of luminosity as a function of halo mass.

For simplicity, we shall assume that all haloes have a NFW structure,
where the local logarithmic slope of the density profile tends
asymptotically to $-1$ for small radii, even
though the Aquarius project, a recent set of simulations aimed at following
the formation of MW-sized haloes at different mass resolutions, demonstrated
that the Einasto profile provides an even better fit \citep{Navarro-08}, a
finding that is also supported by the independent `GHALO' simulation
\citep{Stadel-09}. In the Einasto profile, the local logarithmic slope of
the density profile changes with radius according to a power law with
exponent $\alpha_{sh}$ (typically called shape parameter). According to the
Aquarius project, $\alpha_{sh}$ is of the order $0.16-0.17$ and varies from halo to
halo \citep{Navarro-08}. Previously, \citet{Gao-08}
found a dependence on halo mass for the value of the shape parameter
(going from $0.16$ for galaxy-haloes to $0.3$ for massive clusters).
However, for the purposes of our work, the NFW profile is
a good enough approximation and using it simplifies comparisons to a number of
results in the literature. In any case, we note that using the Einasto profile
(with a typical shape parameter of 0.17)
instead would increase our results for the fluxes by $\sim 50\%$. 
This value changes only very slightly over the range of values for
$\alpha_{sh}$ found for dark matter haloes. Such 
an uncertainty on the dark matter density profile is 
considerably lower than others in our analysis, namely,
the photon spectrum from annihilation (see Fig.~\ref{susy_energy}) and the
contribution from unresolved haloes and subhaloes (see sections 5.3 and 5.4).

Nevertheless, it is important to take into account the impact of finite
numerical resolution in the measured values of $V_{\rm max}$ and $r_{\rm max}$
for the smallest haloes that can be resolved. As has been shown by
\citet{Springel-08a}, the values of $r_{\rm max}$ are increasingly
overestimated for smaller haloes whereas for $V_{\rm max}$ the opposite holds.
A partial correction for these numerical effects can be obtained by trying to 
account for the
effect of the gravitational softening $\epsilon$ for haloes with values of
$r_{\rm max}$ close to $\epsilon$. Assuming adiabatic invariance of orbits
between softened and unsoftened versions of the same halo,
one obtains to leading order:
\begin{eqnarray}\label{soft_cor}
r_{\rm max}'=r_{\rm max}[1+(\epsilon/r_{\rm max})^2]^{-1}\nonumber \\
V_{\rm max}'=V_{\rm max}[1+(\epsilon/r_{\rm max})^2]^{1/2} ,
\end{eqnarray}
where the primed quantities refer to the case without softening.

\begin{figure}
\centering
\includegraphics[height=10cm,width=8.8cm]{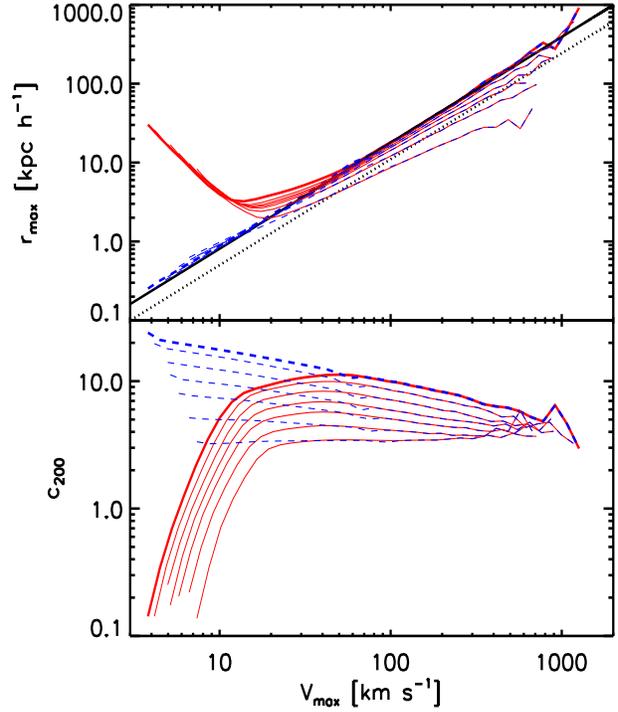}
\caption{{\it Upper panel:} Relation between $r_{\rm max}$ and $V_{\rm
    max}$ for all main haloes in the MS-II at different redshifts. The
  red solid lines are based on measurements corrected for softening effects
  with Eqs.~(\ref{soft_cor}). The blue dashed lines show the values we adopt
  for calculating the expected luminosity of a halo; here an
  additional correction for $r_{\rm max}$ was applied for small haloes
  such that they extend the power law behaviour measured for well
  resolved structures. The black solid line is an extrapolation of the
  results found by \citet{Neto-07} for the MS-I, while the black
  dotted line is the power law obeyed by subhaloes, see
  Fig.~\ref{Vmax_rmax_subhaloes}.  The thicker lines correspond to
  $z=0$, while the rest of the lines are for redshifts: 0.5, 1.0, 1.5,
  2.1, 3.1 and 4.9. {\it Bottom panel:} The relation between
  concentration $c_{200}$ and $V_{\rm max}$. The line styles and
  colors are the same as in the upper panel.}
\label{Vmax_rmax_haloes}
\end{figure}

In general, the value of $r_{\rm max}$ tends to be much more affected
by the softening than $V_{\rm max}$.  This can be clearly seen in the
upper panel of Fig.~\ref{Vmax_rmax_haloes} where we show the $V_{\rm
  max}$-$r_{\rm max}$ relation for all main haloes of the MS-II. Note
that the relation has some intrinsic scatter, but we here only show
the mean values of $r_{\rm max}$ in logarithmic bins of $V_{\rm
  max}$. The red solid lines show the correlation for different redshifts
equal to $z=0.5$, $1.0$, $1.5$ , $2.1$, $3.1$, and $4.9$, after the
softening correction of Eqs.~(\ref{soft_cor}) has been applied,
whereas the thicker line is for $z=0$. The black solid line is an
extrapolation of the results found by \citet{Neto-07} at $z=0$ for the
MS-I.  There is a clear upwards deviation from the power-law behaviour
for haloes with particle number below $N_p\sim 3600$ (corresponding to
a mass of $\sim 2.5\times10^{10}\hMsol$, or $V_{\rm max}\sim 60\kms$ at
$z=0$). This deviation indicates that the softening correction is only
able to correct part of the resolution effects.  We can be certain
that this is indeed a numerical effect by looking at the corresponding
results for the Aquarius haloes \citep{Springel-08a}, which
demonstrate with simulations of the same initial conditions but
drastically different mass resolutions that the $r_{\rm max}$-$V_{\rm
  max}$ relation is indeed a power law over a large dynamic range, at
least for $V_{\rm max}$ as small as $\sim1.5\kms$. This is the case 
for both, main haloes and subhaloes (see Fig. 26 of
\citet{Springel-08a}). We note that haloes with $N_p\sim1000$ or less are 
affected importantly by two body relaxation. The discrete representation of 
a dark matter halo with this low number of particles will undergo two-body 
encounters whereas a real dark matter system with $\sim100~$GeV particles 
is essentially collisionless. Generically, the two-body relaxation 
process tends to reduce the concentration of objects \citep[e.g.][]{Diemand-04}. 
We think this is the main mechanism responsible for the 
upturn shown in Figs.~\ref{Vmax_rmax_haloes} and \ref{Vmax_rmax_subhaloes},
even though a detailed study of this effect is out of the scope of this work.

Since an overestimate of $r_{\rm max}$ can result in an underestimate of the
computed luminosity for a given halo, we further correct the $r_{\rm max}$
values of small haloes phenomenologically in order to allow use of all haloes
in the MS-II down to the minimum resolved halo mass of $1.4\times10^8\hMsol$,
and to get the correct mean halo luminosity over the full mass range.  For
this purpose, we force the values of $r_{\rm max}$ for all haloes with less
than $3600$ particles to follow the power law
\begin{equation}\label{power_law_haloes}
r_{\rm max}''=A(z)\left(\frac{V_{\rm max}'}{{\rm km\,s^{-1}}}\right)^{\alpha(z)}
\end{equation}
where $A(z)$ and $\alpha(z)$ are the parameters of the power law.  These
parameters actually depend very mildly on redshift (with $A(0)\approx0.04$ and
$\alpha(0)\approx1.32$ at $z=0$). The redshift dependence can be understood as
a change in the mass-concentration relation of haloes, which gets flatter for
increasing redshift \citep{Zhao-03,Gao-08}. We estimate the concentration of the main
haloes following the description given in \citet{Springel-08a}, where a
measure of the concentration is obtained in terms of
the mean overdensity within
$r_{\rm max}$ relative to  the critical density $\rho_{\rm crit}$:
\begin{equation}\label{delta_V}
  \delta_V=\frac{\bar{\rho}(r_{\rm max})}{\rho_{\rm crit}}=2\left(\frac{V_{\rm
        max}}{H(z)r_{\rm max}}\right)^2 ,
\end{equation}
where $H(z)$ is the Hubble rate as a function of redshift.  The overdensity
$\delta_V$ can be related to the usual concentration parameter
$c_{200}=r_{200}/r_s$ of the NFW profile using the characteristic overdensity:
\begin{equation}\label{delta_c}
\delta_c=\frac{200}{3}\frac{c_{200}^3}{\ln(1+c_{200})-c/(1+c_{200})}
=\frac{\rho_s}{\rho_{\rm crit}}=7.213\,\delta_V .
\end{equation}
Using Eq.~(\ref{delta_c}) we can compute $c_{200}$ for a given value of
$V_{\rm max}$ and $r_{\rm max}$. The lower panel of
Fig.~\ref{Vmax_rmax_haloes} shows the relation between $c_{200}$ and $V_{\rm
  max}$ for the corrected and uncorrected values of $r_{\rm max}$ as blue
dashed and red solid lines, respectively, with the redshift increasing from top to bottom. The
change in slope of the power law followed by the blue dashed lines is roughly in agreement
with the results of \citet{Zhao-03} and \citet{Gao-08}.

\subsection{Gamma-ray luminosity of haloes down to the damping scale limit}

Following the formulation in section 2, we here analyze the flux multiplier
for large volumes, Eqs. ~(\ref{flux_mmin}) and (\ref{flux_mmin_2}), for the
MS-II. Recall, the flux multiplier gives the ratio of the $\gamma$-ray flux
coming from all haloes inside the simulated volume with masses larger than a
minimum mass $M_{\rm min}$ to the emission produced by a homogeneous
distribution of dark matter filling the box of volume $V_B$ with an average
density $\bar{\rho}_B$.

\begin{figure}
\centering
\includegraphics[height=8.5cm,width=8.5cm]{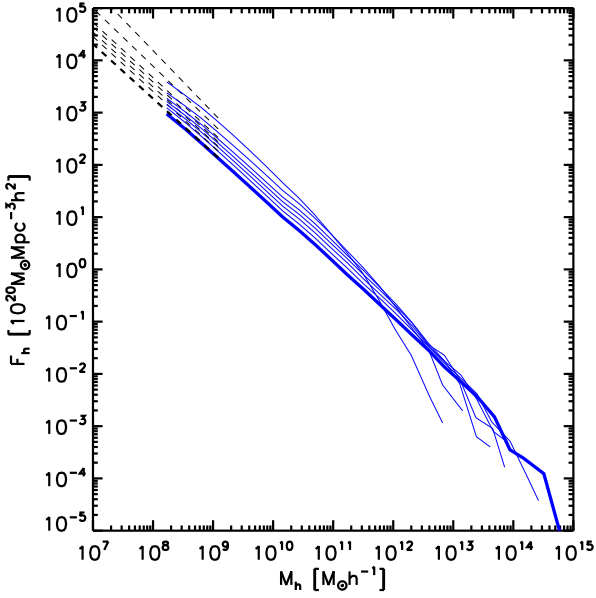}
\caption{Total luminosity coming from main haloes per differential
  mass interval as a function of mass.  The thin blue lines are for
  different redshifts as in Fig.~\ref{Vmax_rmax_haloes}, while the
  thick solid line shows the $z=0$ result.  The dashed lines show the
  run of our extrapolations as discussed in the text.}
\label{diff_Lum}
\end{figure}

We obtain this dimensionless flux multiplier by defining first the function
\begin{equation}\label{F_M}
F_{h}(M_{h})=\frac{\sum L_{h}}{\bar{M}_h\Delta \log M_{h}},
\end{equation}
where the sum is over all the luminosities $L_h$ of haloes with masses in the
logarithmic mass range: $\log M_{h}\pm \Delta \log M_{h}/2$, where $\Delta
\log M_{h}$ is a fixed logarithmic bin size; $\bar{M}_{h}$ is the mean value
of the halo mass in the given bin. Using this definition we can approximate
Eq.~(\ref{flux_mmin_2}) as
\begin{equation}\label{flux_mmin_sim}
  f(M_h>M_{\rm min})\sim\frac{1}{\bar{\rho}_B^2V_B}\int_{M_{\rm
      min}}^{\infty}\frac{F_h(M_h)}{\ln 10}\,{\rm d} M_h .
\end{equation}
In this sense, the function $F_h(M_h)$ is just the total luminosity of haloes
in a mass range, per unit mass range. The function $F_h(M_h)$ is shown in
Fig.~\ref{diff_Lum} for all main haloes in the MS-II. The different blue lines
are for different redshifts, as in Fig.~\ref{Vmax_rmax_haloes}.

At intermediate mass ranges, $F_h(M_h)$ is clearly well approximated by a
power law:
\begin{equation}\label{F_M_power}
F_h(M_h,z)=A_h(z)M_h^{\alpha_h(z)}.
\end{equation}
Our goal is to fit the parameters of this power law so that an extrapolation
can be done down to the cutoff mass for neutralinos. For the
neutralino mass corresponding to the model we have chosen, the free
streaming mass is of the order of $10^{-7}\hMsol$
\citep{Hofmann-Schwarz-Stocker-01}, however, acoustic oscillations due to the
coupling between cold dark matter and the radiation field in the early Universe,
can also produce a damping in the power spectrum of density perturbations
\citep[e.g.][]{Loeb-Zaldarriaga-05}. The cutoff mass of the smallest haloes
that can be formed is determined by the strongest of these
effects. Taking the recent results of \citet{Bringmann-09} (see their
Fig.~3), this cutoff mass for $m_{\chi}=185$ GeV lies in the range $10^{-9}-10^{-4}{\rm
  M}_\odot$. We will take a fiducial value of $10^{-6}\hMsol$ for our
extrapolation, noting that the value of the minimum mass for bound
neutralino dark matter haloes is a source of uncertainty in our results.

We obtain the parameters of the power law in Eq.~(\ref{F_M_power}) by fitting 
the function $F_h(M_h)$ between two mass limits, with the lower limit chosen as $M_{\rm
  lim,min}=6.89\times10^8\hMsol$, corresponding to haloes with 100 particles
(below this number the mass and abundance of haloes is not reliable), and the
higher limit set equal to the last logarithmic mass bin with more than 500
haloes, such that uncertainties from counting statistics are avoided. We find
that for these mass ranges, the parameters of the power law fits change only
slightly with redshift; in fact for $z<2.1$, $\alpha_h\simeq -1.05$ with less
than $2\%$ variation, and $A_h\simeq 6.92\times10^{11}$ with less than
$50\%$ variation. The black dashed lines in Fig.~\ref{diff_Lum} show the
resulting extrapolation of the power law down to $10^7\hMsol$.

\begin{figure}
\centering
\includegraphics[height=8.5cm,width=8.5cm]{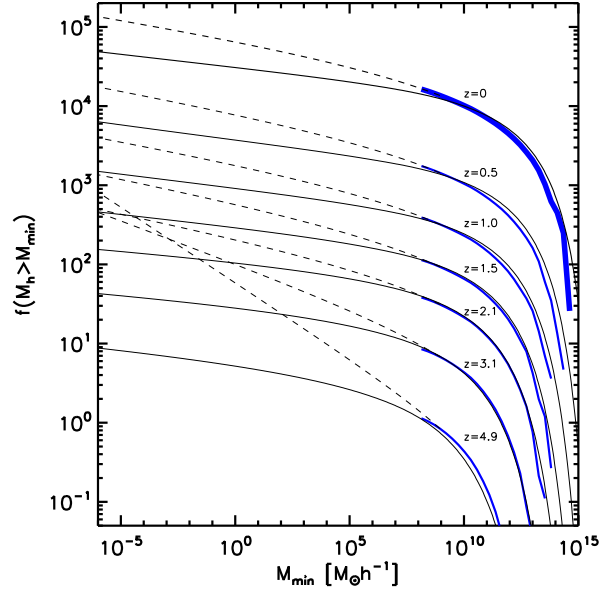}
\caption{Flux multiplier $f(M_{\rm h}>M_{\rm min})$ for the main haloes in the
  MS-II as a function of $M_{\rm min}$. The solid-blue and
  dashed-black lines are analogous to the ones in
  Fig.~\ref{diff_Lum}. The solid-black line is a theoretical
  estimate as described in the text.}
\label{Lum_cumulative}
\end{figure}

The blue lines in Fig.~\ref{Lum_cumulative} give results for the flux
multiplier, Eq.~(\ref{flux_mmin_sim}), as a function of $M_{\rm min}$ for the
main haloes in the MS-II for different redshifts, as in in
Fig.~\ref{diff_Lum}.  The black dashed lines show the extrapolation of the
flux multiplier down to $M_{\rm min}=10^{-6}\hMsol$ using
Eq.~(\ref{F_M_power}). The flux multiplier has been previously computed
analytically by \citet{Taylor-Silk-03}.  We follow their procedure and
calculate an analytical estimate of $f(M>M_{\rm min})$ for the cosmology
corresponding to the MS-II. We first use Eq.~(\ref{flux_nfw}) to compute the
individual flux multiplier $f_h(c_{\Delta})$ of an NFW halo with a
concentration $c_{\Delta}$ given by the redshift mass-concentration relation,
computed as in the analytical model of \citet{Eke-Navarro-Steinmetz-01} and
with a choice of $\Delta=178\, \Omega_m^{0.45}\sim95$.  Using the value of
$f_h(c_{\Delta})$ we then compute the full flux multiplier based on
Eq.~(\ref{flux_mmin_2}), assuming a mass function calculated with the
formalism of \citet{Sheth-Mo-Tormen-01}.  The result is included in
Fig.~\ref{Lum_cumulative} with black solid lines.

We see that these analytic estimates roughly agree with the actual numerical
results from the MS-II over the resolved mass range.  However, small
differences in the slope of $f(M>M_{\rm min})$ at the minimum resolved mass
range ($1.39-6.89\times10^8\hMsol$), produce a large difference for the
extrapolated values at the damping scale limit, which amounts to a factor of up
to $ 3-5$. We note that for redshifts $z>2.1$, our extrapolation down to the
cutoff mass is not accurate any more and the obtained values are
clearly overestimated. This is a reflection of the difficulties to still
reliably fit a power law to $F(M)$ at these high redshifts, where the
population of haloes over the resolved mass range becomes ever smaller.

\begin{figure}
\centering
\includegraphics[height=8.5cm,width=8.5cm]{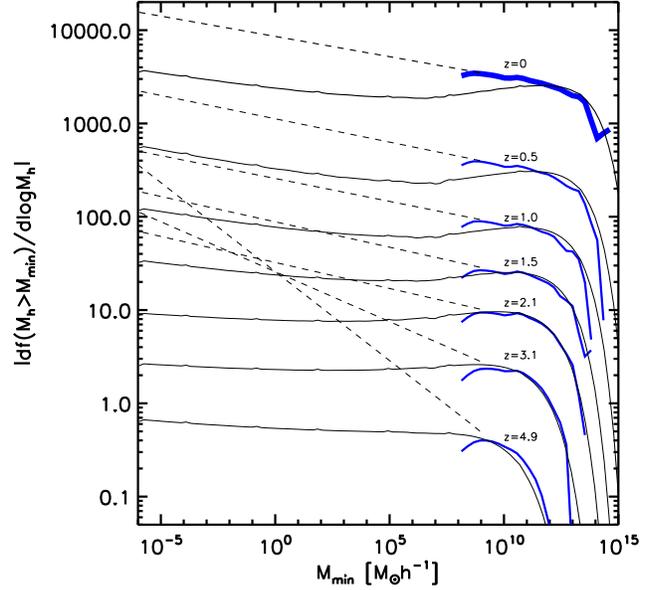}
\caption{Differential cumulative flux multiplier $\vert{\rm d}f(M_{\rm
    h}>M_{\rm min})/{\rm d}\log M_{\rm min}\vert$ as a function of
  minimum mass. The line styles and colors are as in
  Fig.~\ref{Lum_cumulative}.}
\label{Lum_cumulative_diff}
\end{figure}

Another interesting question that arises from this statistical analysis
is: Which mass scale contributes most to the total flux of gamma rays
coming from dark matter annihilations?  The answer is presented in
Fig.~\ref{Lum_cumulative_diff}, which shows the differential cumulative flux,
which is really just the logarithmic derivative of the cumulative flux shown
in Fig.~\ref{Lum_cumulative}, as a function of the minimum halo mass. From the
figure we see that less massive haloes contribute increasingly more to the
total gamma-ray flux coming from dark matter annihilation. There is a clear
change in behaviour for a given mass scale $M_{\rm min}\sim10^{14}\hMsol$ at
$z=0$, going down to $M_{\rm min}\sim10^{12}\hMsol$ at
$z=2.1$. Fig.~\ref{Lum_cumulative_diff} also reveals an artificial downturn
for the minimum mass range resolved ($1.39-6.89\times10^8\hMsol$). The
analytical expectations, shown as black solid lines, qualitatively agree with
the results from the simulation. However, the quantitative differences in the
extrapolated region are large.

\subsection{The effects of substructure}

Substructures within CDM haloes produce a total $\gamma$-ray luminosity which
is dominant over the smooth halo component for an external observer
\citep[e.g][]{Taylor-Silk-03}. \citet{Springel-08b} showed that for a MW halo,
substructures with masses larger than $10^5$M$_{\odot}$ within $r_{200}$ emit
as much as $76\%$ of the emission of the smooth halo. Also, using their set of
simulations of the same halo at different resolutions they extrapolated the
contribution from substructures down to masses of $10^{-6}$M$_{\odot}$, at
which point they found the total cumulative luminosity to be 232 times larger
than the contribution from the smooth halo component.

In this subsection, we analyse the contribution of substructures to the
$\gamma$-ray flux of the main haloes using as a basis the most massive haloes
in the MS-II that contain significant resolved subhaloes and are therefore
suitable for such an analysis. To this end we follow an analogous procedure to
the one used in the previous subsection.

\begin{figure}
\centering
\includegraphics[height=8.5cm,width=8.5cm]{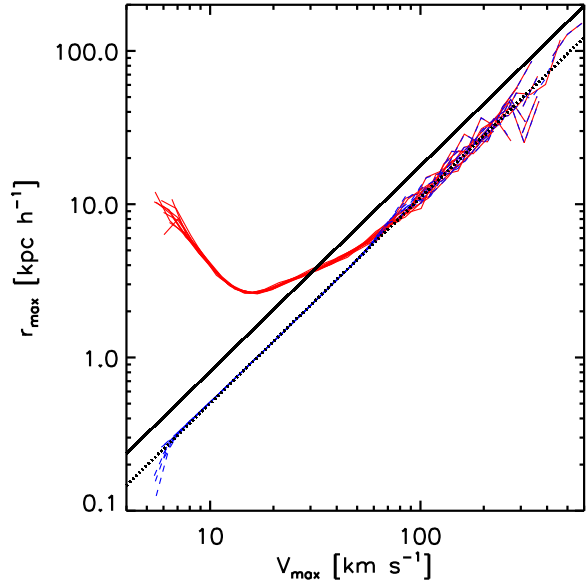}
\caption{Relation between $r_{\rm max}$ and $V_{\rm max}$ for the
  subhaloes of the 10 most massive clusters in the MS-II at $z=0$. These
  main haloes have $V_{max}$ values in the range $\sim850-1250$kms$^{-1}$.
  Line styles and colors are as in Fig.~\ref{Vmax_rmax_haloes}}
\label{Vmax_rmax_subhaloes}
\end{figure}

Fig.~\ref{Vmax_rmax_subhaloes} shows the $r_{\rm max}-V_{\rm max}$
relation for the population of subhaloes inside the 10 most massive
clusters of the MS-II at $z=0$, the line styles and colors are the
same as in Fig.~\ref{Vmax_rmax_haloes}. These clusters have $V_{max}$
values in the range $850-1250$~kms$^{-1}$. The subhaloes follow a very
similar power-law behaviour as the population of main haloes in the
upper panel of Fig.~\ref{Vmax_rmax_haloes}. The slope of the power law
is the same, only the normalization is slightly different, a factor of
$0.62$ lower than for main haloes. This is in agreement with findings
obtained for the Aquarius project \citep{Springel-08a}.

As for main haloes, we have corrected the values of $r_{\rm max}$ and
$V_{\rm max}$ of the subhaloes for softening effects according to
Eqs.~(\ref{soft_cor}). Also, as for the main haloes, the upturn in the
$r_{\rm max}-V_{\rm max}$ relation for small subhaloes clearly
indicates an overestimation of $r_{\rm max}$ in this regime due to
numerical resolution effects. In order to avoid biasing our results,
we hence force the subhaloes in this low-$r_{\rm max}$ low-$V_{\rm
  max}$ regime to continue to follow the power law defined by the
larger, more massive subhaloes:
\begin{equation}\label{power_law_subs}
r_{\rm max}''=A_{\rm sub}(z)(V_{\rm max}')^{\alpha_{\rm sub}},
\end{equation}
where $\alpha_{\rm sub}$ has a slight redshift dependence that we can
neglect since a value of $\alpha_{\rm sub}\sim1.34$ fits the mean
behaviour of the data for $z<3$ with an accuracy better than $20\%$.
We find that the value of $A_{\rm sub}(z)$ can be approximated as:
\begin{equation}\label{A_subs}
A_{\rm sub}(z)\sim0.023(1+z)^{-1/2}.
\end{equation}
As a further consistency check, we have analyzed all main haloes in
the MS-II with more than $500$ subhaloes, finding that they agree well
with the behaviour of the 10 most massive clusters shown in
Fig.~\ref{Vmax_rmax_subhaloes}.  Thus, all resolved subhaloes in the
MS-II with particle number $N_{\rm p}<3600$ were corrected using
Eq.~(\ref{power_law_subs}).

The contribution of subhaloes, per unit mass range, to the total
$\gamma$-ray luminosity of their host can be analyzed using the
following quantity:
\begin{equation}\label{F_subs}
F_{\rm sub}\left(\frac{M_{\rm sub}}{M_{h}}\right)=\left(\frac{M_{h}}{L_{h}}\right)\frac{\sum
L_{\rm sub}}{\bar{M}_{\rm sub}\Delta \log M_{\rm sub}}
\end{equation}
where $M_{\rm sub}$ and $L_{\rm sub}$ are the mass and gamma-ray
luminosity of a given subhalo. We have computed $F_{\rm sub}$ for all
main haloes in the MS-II with more than $500$ subhaloes and found
that this quantity can be described by a power law in the intermediate
mass range, in a similar way to the function $F_h$ for main haloes:
\begin{equation}\label{F_subs_power_law}
F_{\rm sub}\left(\frac{M_{\rm sub}}{M_{h}}\right)\sim
A_{\rm sub}\left(\frac{M_{\rm sub}}{M_{h}}\right)^{\alpha_{\rm sub}}
\end{equation}
We estimate the parameters of this power law by fitting the results
between $M_{\rm sub,min}=6.89\times10^8\hMsol$, the same lower mass
limit used for main haloes, and an upper mass limit given by the most
massive logarithmic mass bin with more than 15 subhaloes. Using these
two mass limits a power law is found to provide a good fit to the
general behaviour in the intermediate mass range for all main halo
masses and at different redshifts.

\begin{figure}
\centering
\includegraphics[height=10cm,width=8.8cm]{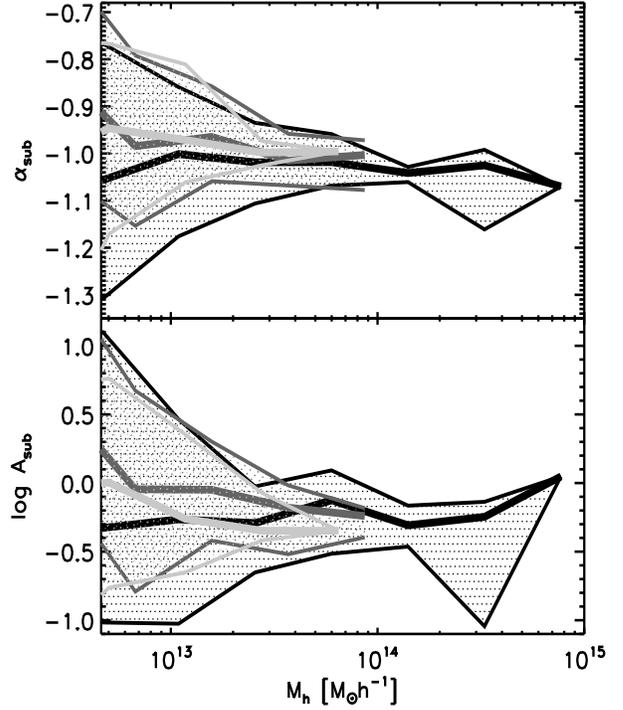}
\caption{Parameters $\alpha_{\rm sub}$ and $A_{\rm sub}$ from
  Eq.~(\ref{F_subs_power_law}) as a function of main halo mass.  The
  thickest solid lines are the median values for logarithmic mass bins, the
  dotted regions are within the first and third quartiles.  The black, dark grey and
  light grey lines are for $z=0$, $1.0$ and $2.0$, respectively.}
\label{subs_power_param}
\end{figure}

Fig.~\ref{subs_power_param} shows the parameters $A_{\rm sub}$ and
$\alpha_{\rm sub}$ of the fit as a function of the mass of the
host. The median values for logarithmic mass bins at $z=0$ are shown
with thick black solid lines, the first and third quartiles with thiner black
solid lines. The dark grey and light grey lines are for $z=1.0$ and $z=2.0$,
respectively.  The results show almost no dependence of the parameters
of the power law fit on the mass of the host and on redshift. For the
majority of the cases, these parameters lie in the range:
\begin{eqnarray}\label{param_range}
-0.95\leq\alpha_{\rm sub}\leq-1.15\nonumber\\ -0.5\leq \log A_{\rm sub}
\leq 0.1 .
\end{eqnarray}
For lower host halo masses, the scatter in the parameters for a given
mass bin becomes increasingly larger, however. In these cases, the low
number of substructures per halo makes the power-law fit less
reliable.

We will assume that the power law in Eq.~(\ref{F_subs_power_law}) is
universal, allowing us to extrapolate the value of $F_{\rm sub}$ down to the
damping scale limit. Using the range in Eqs.~(\ref{param_range}), we can
obtain estimates for the
maximum and minimum contribution of all
substructures to the gamma-ray luminosity of their host halo:
\begin{equation}\label{subs_extrapol}
f_{\rm sub}(f_{\rm max},M_h)\sim\frac{1}{L_h}\int_{10^{-6}}^{f_{\rm max}M_h}\left(\frac{L_h}{M_h}\right)
\frac{F_{\rm sub}\left(\frac{M_{\rm sub}}{M_{h}}\right)}{\ln 10}{\rm
  d}M_{\rm sub} ,
\end{equation}
where $f_{\rm max}$ is the ratio of the most massive subhalo to the
mass of the host. If $\alpha_{\rm sub}\neq-1$ then:
\begin{equation}\label{subs_extrapol_1}
f_{\rm sub}(f_{\rm max},M_h)\sim\frac{A_{\rm sub}\left[ (f_{\rm
      max})^{\alpha_{\rm
        sub}+1}-\left(\frac{10^{-6}}{M_h}\right)^{\alpha_{\rm
        sub}+1}\right]}{\ln 10(\alpha_{\rm sub}+1)}.
\end{equation}
For the range of values given in Eqs.~(\ref{param_range}), which
reflects the uncertainties in our power law extrapolation, the total
luminosity of all substructures in a halo of mass $10^{12}\hMsol$ lies
in the range $f_{\rm sub}\in(2,1821)$ times the luminosity of the main
halo, bracketing the value of 232 found by \citet{Springel-08b} for
the Aq-A halo of the Aquarius project.

\section{Contributions to the EGB from dark matter annihilation}

With the results of the previous sections, we finally concentrate on
the main objective of this work, namely, to calculate the contribution
of dark matter annihilation to the EGB. Specifically, we are
interested in making realistic maps of the expected specific intensity
across the sky.

\subsection{Simulation of the past light cone}

The specific intensity (Eq.~\ref{intensity}) describes the total
emission from dark matter annihilations integrated over the full
backwards light cone along a certain direction.  We use the data of
the MS-II to fill the whole volume contained in the past-light cone of
an observer located at a fiducial position in the box at $z=0$. As we
have 68 simulation outputs in total, we can approximate the temporal
evolution of structure growth by using at each redshift along the past
light-cone the output time closest to this epoch. To cover all space,
we use periodic replication of the simulation box. 

\begin{figure}
\centering
\includegraphics[height=7cm,width=7cm]{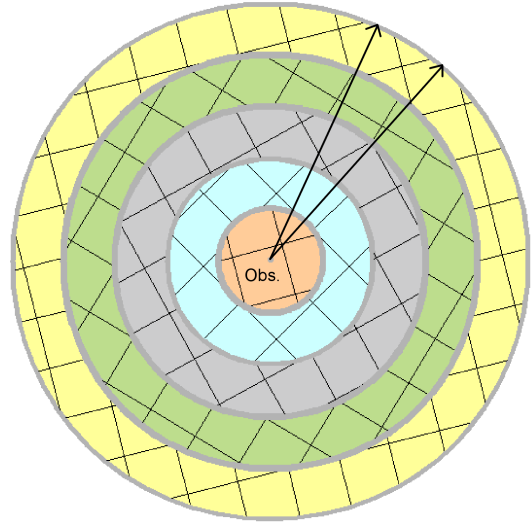}
\caption{Sketch of the light-cone reconstruction carried out with the
  periodic MS-II simulation. We subdivide the light-cone into shells
  of comoving thickness $100\,h^{-1}{\rm Mpc}$. Within each shell, a
  coherent randomization of the orientation and translation of the
  periodic tessellation with the simulation box is carried out, in
  order to avoid repetition of the same structures along a given
  line-of-sight and at the same time to maintain continuity in the
  transverse direction across the sky. The time evolution of cosmic
  structure is taken into account by using at each position (redshift)
  along a line-of-sight the closest simulation output time.}
\label{light_cone}
\end{figure}

However, this periodic replication would introduce strong correlations
along the line-of-sight on the scale of the box size, and in
particular, would lead to replications of the same structures along
certain sight lines. To avoid this problem, we subdivide the backwards
light-cone into shells of comoving thickness equal to the
boxsize. Within each shell, an independent random rotation and
translation is applied to the pattern of boxes that tessellates the
shell, as sketched in Fig.~\ref{light_cone}. This procedure, first
used in \citet{Carbone-08}, eliminates the unwanted line-of-sight
correlations, while at the same time it maintains continuity of the
maps in the transverse direction across the sky. As a result, our
final maps are free of obvious tessellation artifacts.  We note,
however, that this procedure can not make up for the missing large
scale power on scales larger than the box size itself. Also, it will
not be able to eliminate potential excess power in the angular power
spectrum on the periodicity scale of the box, as within each shell the
simulation box seamlessly tiles the shell. However, the fixed comoving
box size translates to a different angular scale in each shell, so
that the averaging over many shells usually makes this a small effect.

As discussed earlier, our strategy is not to account for the emission
at the level of individual particles, but instead to use entire haloes
and subhaloes, allowing us to accurately correct for resolution
effects. The list of dark matter particles at each output time is
hence replaced by a catalogue of dark matter substructures, each with
a known total luminosity and known radial luminosity profile,
calculated from their $r_{\rm max}$ and $V_{\rm max}$ values as
described earlier, and including a resolution correction for poorly
resolved objects. Map-making then becomes a task to accurately
accumulate the properly redshifted emission from these structures in
discretized representations of the sky. For the pixelization of the
sky we use equal area pixels based on the {\small
  HEALPIX}\footnote{The HEALPIX software package is available at
  http://healpix.jpl.nasa.gov} tessellation \citep{Gorski-05}.  All
our maps use $N_{\rm pix}=12\times 512^2\sim \pi \times 10^6$ pixels
corresponding to an angular resolution of $\sim 0.115\degr$.

A given pixel in the simulated maps covers a solid angle
$\Delta\Omega_{\rm pix}$. Our map-making code computes the average
value of the specific intensity within the area subtended by this
solid angle by conservatively distributing the emission of each
substructure over the appropriate pixels.  Combining
Eqs.~(\ref{emiss}) and (\ref{intensity}) we can write the relevant sum
over all substructures in the light-cone as:
\begin{equation}\label{intensity_pix}
I_{\gamma,0}(\Delta\Omega_{\rm pix})=\frac{1}{8\pi}\sum_{h\epsilon\Delta\Omega_{pix}}L_{h}w(d_h,r_h)E_{\gamma,0}f_{\rm
  SUSY}(z_h)\vert_{E_{\gamma,0}}
\end{equation}
where the function $w(d_h,r_h)$ is a weight function that distributes
the luminosity of a given halo onto the pixels overlapping with the
projected ``size'' (or more precisely the luminosity profile) of the
halo; the latter depends on the distance of the observer to the halo
$d_h$ and the transverse distance $r_h$ between the halo centre and
the centre of the pixels it touches. Except for structures that are
very nearby, the high central concentration of the emission of a
subhalo and the limited angular resolution of our maps give most
subhaloes the character of unresolved point sources. It therefore
makes little difference in practice if the transverse luminosity
profile of a subhalo, a $\rho^2$-weighted projection of a NFW profile,
is replaced with an SPH-like kernel function in 2D with radius equal
to the halo's half-mass radius $r_{1/2}$. However, it is important to
guarantee that the sum of the weights $w(d_h,r_h)$ associated with all
pixels that overlap with a given substructure `add up to unity',
i.e.~that the total emission of the halo is represented in the map in
a conservative fashion. Our map-making code guarantees this
property. For performance reasons, the code also uses a tree-based
approach to quickly find all subhaloes (and also any of their periodic
images, if appropriate) that overlap with a given line of sight.

For the purpose of extending the predictions of our maps down to the
damping scale limit of neutralinos ($\sim10^{-6}$M$_{\odot}$), we will
incorporate the results of section 4 by dividing the maps into
separate components: (i) the contribution of {\em resolved haloes and
  subhaloes} with a minimum mass of $6.89\times10^8\hMsol$ (called
``ReHS'' in the following); (ii) the contribution of {\em unresolved
  main haloes} with masses in the range $1.0\times
10^{-6}-6.89\times10^8\hMsol$ (referred to as ``UnH''); and finally,
(iii) the contribution of {\em unresolved subhaloes} in the same mass
range as in the case (ii) (component ``UnS'').

\subsection{Resolved haloes and subhaloes (ReHS)}

\begin{figure*}
\centering
\resizebox{16cm}{!}{\includegraphics{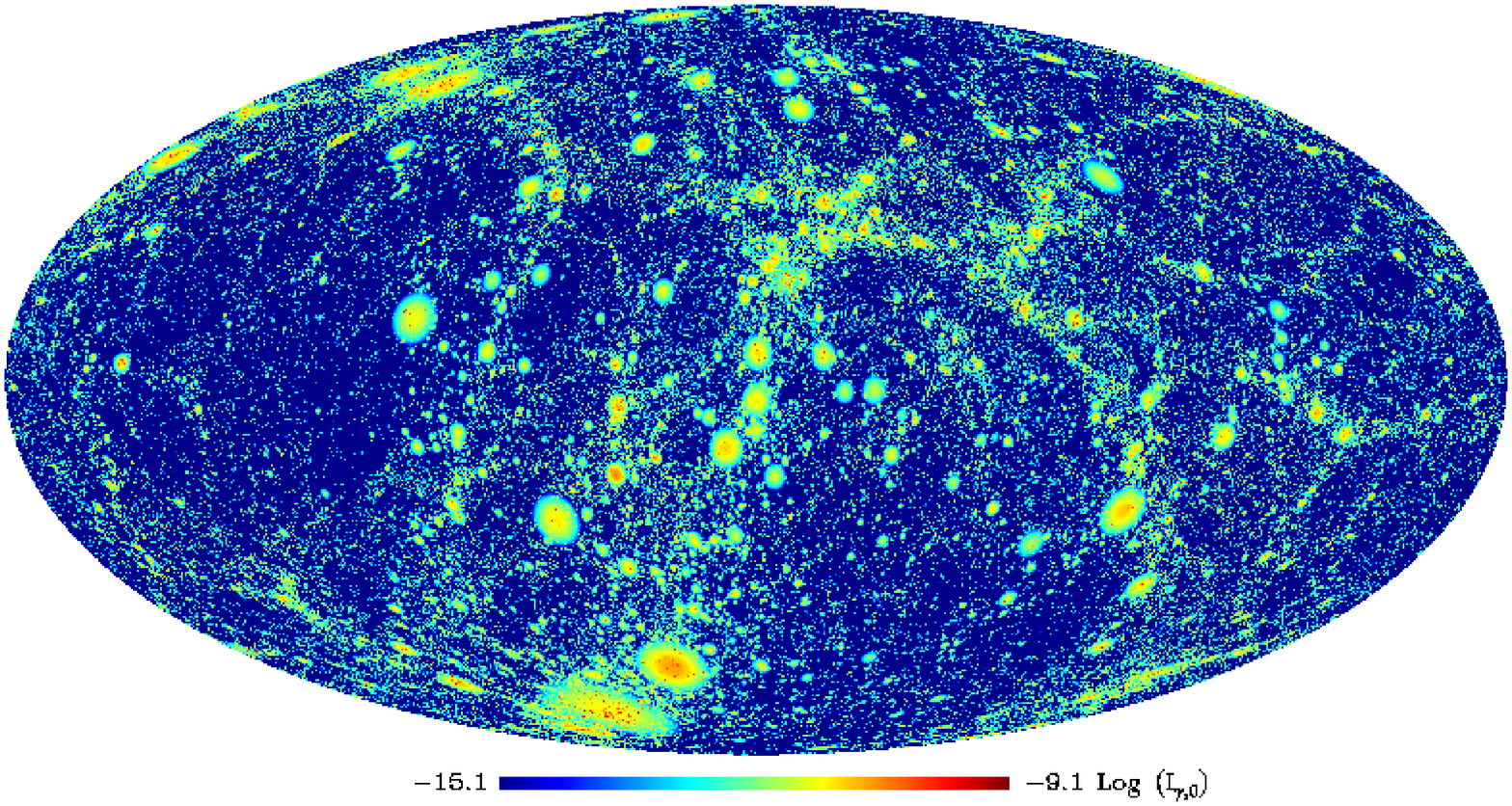}}
\resizebox{16cm}{!}{\includegraphics{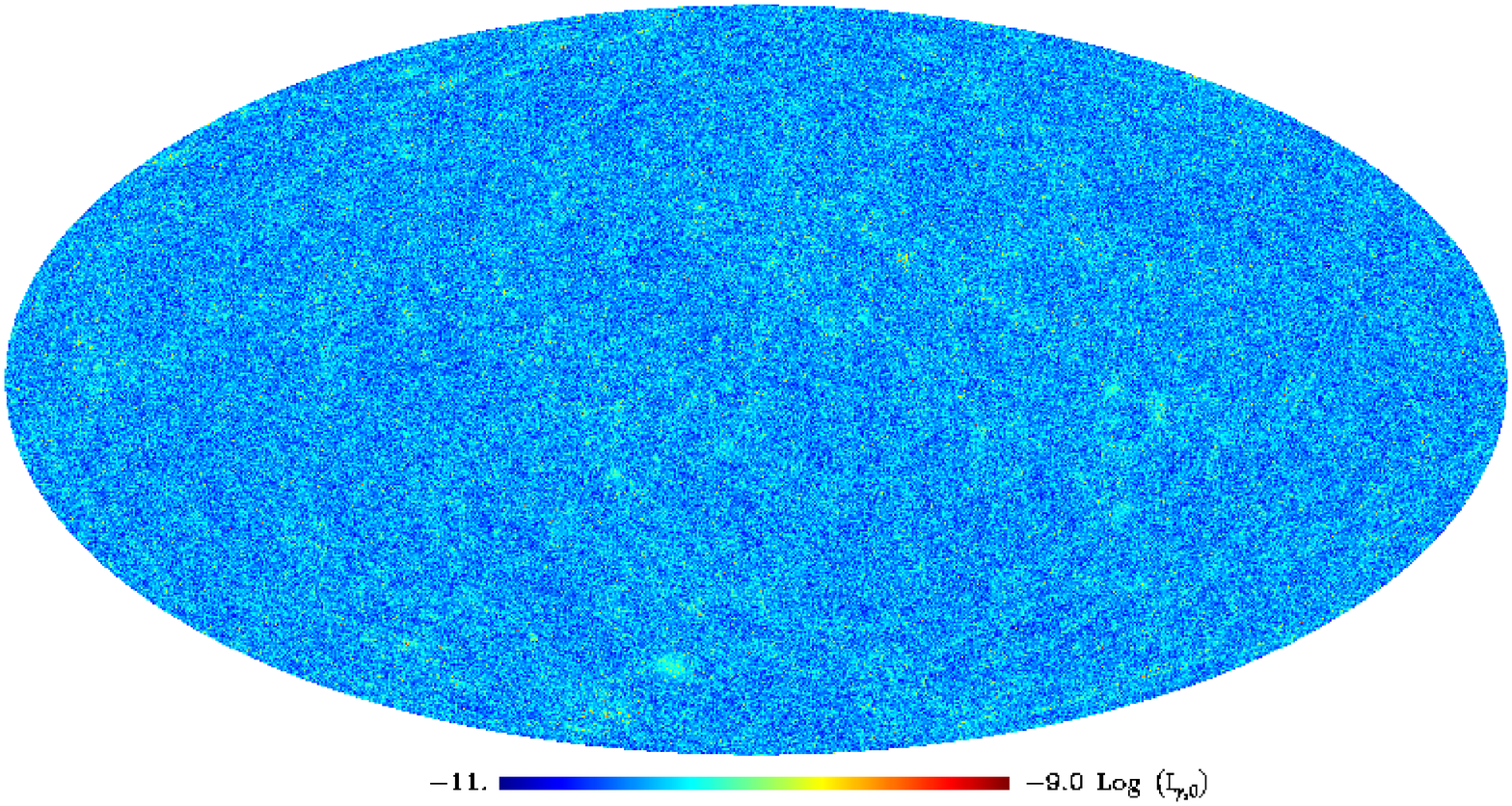}}
\caption{{\it Upper panel:} One of the partial maps ($z=0$) showing
  the cosmic $\gamma$-ray background produced by dark matter
  annihilation.  The color scale gives a visual impression of the
  values of the specific intensity for each pixel in the map; the red
  color corresponds to the highest values of specific intensity.  The
  observed energy of the simulated $\gamma$-ray radiation is $10\,{\rm
    GeV}$, and the benchmark point L as described on Table 1 was used
  as input for the supersymmetric model.  {\it Lower panel:} Co-added
  map showing the full $\gamma$-ray sky map from dark matter
  annihilation integrated out to $z=10$.}
\label{map}
\end{figure*}

In Fig.~\ref{map}, we show full sky maps of the the $\gamma$-ray emission of
all main haloes and subhaloes that are detected in the MS-II; this corresponds
to an essentially perfectly complete sample of all haloes above a mass limit of
$6.89\times10^8\hMsol$. In the top panel of Fig.~\ref{map}, a partial map at
low redshift (corresponding to the first shell at $z=0$ of depth
$50\,h^{-1}{\rm Mpc}$) is shown in order to illustrate typical foreground
structure, whereas the bottom panel gives an integrated map out to $z=10$,
which is approximately the full EGB from annihilation.  The maps in
Fig.~\ref{map} are for $E_{\gamma,0}=10\,{\rm GeV}$, where $E_{\gamma,0}$ is the
photon energy measured by the observer (i.e.~the emission energy of the
photons is $E_{\gamma}=(1+z)E_{\gamma,0}$). In the color scale used for the
maps, red corresponds to the highest and black to the lowest values of the
specific intensity, respectively.

\subsection{Unresolved haloes (UnH)}

The maps shown in Fig.~\ref{map} are complete down to the minimum mass in the
MS-II that we can trust, $\sim6.89\times10^8\hMsol$. In order to make a
prediction of the full EGB coming from dark matter annihilations, we
extrapolate the $\gamma$-ray flux to account for the contribution of all
missing dark matter haloes down to the cutoff mass $10^{-6}\hMsol$. For
this purpose we use the results found in section 4 on the universal power law
behaviour of the functions $F_h(M_h)$ and $F_{\rm sub}(M_{\rm sub}/M_h)$.

The way we incorporate this contribution in the $\gamma$-ray maps is the
following.  We assume that the EGB radiation from the missing haloes in the
mass range $10^{-6}\hMsol$ to $\sim6.89\times10^8\hMsol$ is distributed on
the sky in the same way as the one from the smallest masses we can resolve in
the simulation, which we adopt as the mass range between
$1.4\times10^{8}\hMsol$ and $\sim6.89\times10^8\hMsol$ (haloes with 20 to 100
particles). This assumption can be justified by the clustering bias of dark
matter haloes that appears to approach an asymptotic constant value for low
halo masses \citep[e.g][]{Boylan-Kolchin-09}.  Hence, we compute the value of
a boost factor $b_h$ with which each resolved halo in the mass range
$1.4-6.89\times10^8\hMsol$ needs to be multiplied such that the luminosity of
the unresolved main haloes is accounted for as well. Using the power law behaviour
of $F_h(M_h)$ and $f(M_h>M_{\rm h,min})$, Eqs.~(\ref{F_M_power}) and
(\ref{flux_mmin_sim}), we find that:
\begin{equation}\label{boost_h}
b_h=\frac{f(10^{-6}\hMsol,6.89\times10^8\hMsol)_a}{f(1.4\times10^{8}\hMsol,6.89\times10^8\hMsol)_{\rm
    sim}}\sim60 ,
\end{equation}
where $b_h$ is given by the ratio of the flux multipliers computed
analytically, between the cutoff mass limit and the 100 particle limit, and
computed within the simulation in the last resolved mass range (20-100
particles). We find that $b_h$ is nearly independent of redshift up to the
highest redshift we can reliable measure the power law $F_h(M_h)$, that is for
$z<2.1$.

\subsection{Unresolved subhaloes (UnS)}

To add the contribution of unresolved substructures to the $\gamma$-ray maps,
we first assume that the identified subhalo population is complete down to a
mass limit of $M_{\rm sub}=6.89\times10^8\hMsol$. Below this mass we compute
luminosities using the formulas presented in section 4.2, separating the way
we add this contribution in two cases.

In the case of resolved main haloes with more than 100 particles, we simply
use Eq.~(\ref{subs_extrapol_1}) with $f_{\rm max}M_h=6.89\times10^8\hMsol$ if
the main halo has subhaloes, and $f_{\rm max}=0.05$ otherwise. For the former
case we distribute the extra luminosity coming from unresolved substructures
among the subhaloes of the main host, assuming in this way that unresolved
subhaloes are distributed in the same way as the resolved ones. For the
latter, the extra luminosity is given to the main background subhalo. We note
that in this case the particular value of $f_{\rm max}$ has little impact in
Eq.~(\ref{subs_extrapol_1}) since $M_h\geq6.89\times10^8\hMsol$ and thus, the
right-hand side of the term in brackets is dominant for $\alpha_{\rm sub}<-1$;
for the extreme value of $\alpha_{\rm sub}=-0.95$, values of $f_{\rm max}$
between 0.005 and 0.5 give variations of $f_{\rm sub}$ smaller than
$3.5\%$. We use $f_{\rm max}=0.05$ since this is a typical value in the MS-II.

For the second case of all main haloes, resolved and unresolved, with masses
less than $6.89\times10^8\hMsol$ we use the following strategy.  Since in the
last subsection we described how to get the total luminosity coming from all
main haloes with masses between the damping scale limit and $6.89\times10^8\hMsol$,
we can compute the boost factor $b_{\rm sub}$ for which this luminosity needs
to be multiplied in order to include the full contribution of all subhaloes of the
main haloes in this mass range:
\begin{equation}\label{boost_subs}
  b_{\rm sub}=\frac{\mathfrak{f}_{\rm
      boost}(10^{-6}\hMsol,6.89\times10^8\hMsol)}{\mathfrak{f}_{\rm no-boost}(10^{-6}\hMsol,6.89\times10^8\hMsol)}
\end{equation}
where $\mathfrak{f}_{\rm no-boost}$ is given by:
\begin{eqnarray}\label{boost_subs_1}
\mathfrak{f}_{\rm no-boost}(10^{-6}\hMsol,6.89\times10^8\hMsol)\approx\nonumber\\
\int_{10^{-6}}^{6.89\times10^8}\frac{F_h(M_h)}{\ln 10}\,{\rm d}M_h
\end{eqnarray}
and $\mathfrak{f}_{\rm boost}$ can be written as:
\begin{eqnarray}\label{boost_subs_2}
\mathfrak{f}_{\rm boost}(10^{-6}\hMsol,6.89\times10^8\hMsol)\approx\nonumber\\
\int_{10^{-6}}^{6.89\times10^8}[1+f_{\rm sub}(f_{\rm
  max},M_h)]\frac{F_h(M_h)}{\ln 10}\,{\rm d}M_h .
\end{eqnarray}
After using Eqs. (\ref{flux_mmin_sim}) and (\ref{F_M_power}) we can write:
\begin{eqnarray}\label{boost_subs_3}
b_{\rm sub}=1+\frac{A_{\rm sub}}{{\rm ln}10(\alpha_{\rm
    sub}+1)}\Bigg\{f_{\rm max}^{\alpha_{\rm sub}+1}-(10^{-6})^{\alpha_{\rm sub}+1}\nonumber\\
\left.\left(\frac{\alpha_h+1}{\alpha_h-\alpha_{\rm
    sub}}\right)\frac{(6.89\times10^8)^{\alpha_h-\alpha_{\rm
        sub}}-(10^{-6})^{\alpha_h-\alpha_{\rm
      sub}}}{(6.89\times10^8)^{\alpha_h+1}-(10^{-6})^{\alpha_h+1}}\right\} 
\end{eqnarray}
We found that $b_{\rm sub}$ is roughly independent of redshift. Using the
values of $\alpha_h=-1.05$, $A_{\rm sub}$, $\alpha_{\rm sub}$ determined
earlier, we obtain $b_{\rm sub}\approx2-60$.  We shall take the extreme values
of this range in the following presentation of our results, as these minimum
and maximum values reflect the uncertainties in our extrapolation and should
bracket the true result.

\subsection{Isotropic and anisotropic components of the EGB from dark matter
  annihilation}

\begin{figure}
\centering
\includegraphics[height=10cm,width=8.8cm]{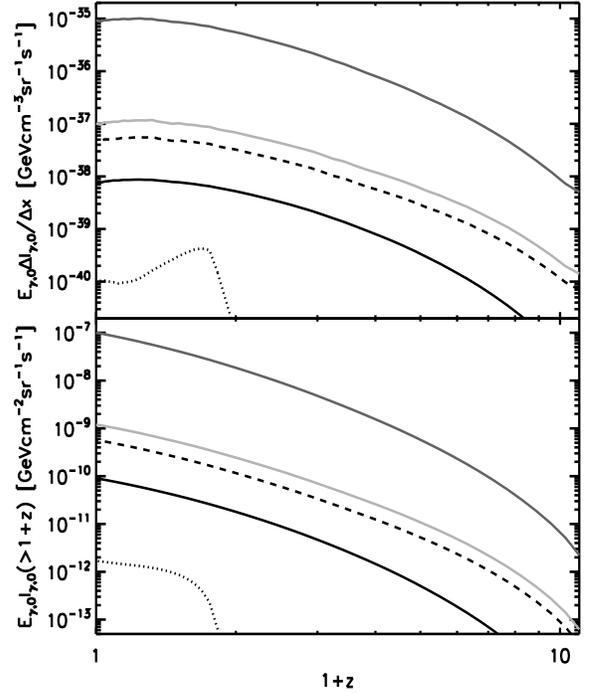}
\caption{{\it Upper panel: }Mean annihilation intensity per comoving
  shell thickness, $\Delta I_{\gamma,0}/\Delta x$, as a function of
  redshift for individual partial maps. The solid black line is for
  resolved haloes and subhaloes. The dashed black line is for main
  haloes only, resolved and unresolved down to the cutoff mass
  using a boost factor $b_h=60$. The light grey and dark grey solid lines show the
  contribution of all components, resolved and unresolved haloes and
  subhaloes, boosted with the extreme values in the interval $b_{\rm
    sub}=2-60$.  All lines are for $E_{\gamma,0}=10\,{\rm GeV}$,
  except for the black dotted line which is for the ReHS component with 
  $E_{\gamma,0}=100\,{\rm GeV}$. {\it Lower panel: } The same as the
  upper panel but for the accumulated intensity,
  $I_{\gamma,0}(>1+z)$.}
\label{maps_means}
\end{figure}

The upper panel of Fig.~\ref{maps_means} shows the mean value of the specific
intensity $\Delta I_{\gamma,0}$ divided by the comoving thickness of the
shell (equal to 100$\hMpc$, except for the map at $z=0$ where the shell is
half as thick) at $E_{\gamma,0}=10\,{\rm GeV}$ for our partial sky
maps as a function of redshift. Recall that each partial map represents the
total specific intensity in a shell of constant comoving thickness, so this
provides information about where the signal is coming from.  The black line is
for the ReHS case, while the dashed line is for main haloes only (resolved and
unresolved, including the boost factor $b_h=60$). The remaining lines are for
the resolved and unresolved main haloes and their subhaloes, boosted to
include the contribution of structures all the way down to the damping scale
limit; here the light grey line is for the ``minimum boost'' with $b_{\rm sub}=2$
whereas the dark grey line is for the ``maximum boost'' with $b_{\rm sub}=60$, but
both include the boost $b_h=60$ for unresolved main haloes.

In all cases, the contribution from partial maps up until $z\sim1$ is almost
constant, indicating that the increasing number of sources for shells at
higher redshifts, due to the larger volume seen behind each pixel, is
approximately compensated by the distance factor and the spectral effects from
the cosmological redshifting and the intrinsic variation of the emission
spectrum, as described by $f_{\rm SUSY}$ (see Fig.~\ref{susy_energy}).

The lower panel of Fig.~\ref{maps_means} shows the accumulated mean value of
the specific intensity as a function of redshift. This clearly shows that the
most relevant contributions come from maps up to $z\sim2$. For higher
redshifts, the partial maps contribute less than $5\%$ to the total. Overall,
applying the maximum boost increases the mean value of $I_{\gamma,0}$ of the
ReHS case by three orders of magnitude.  However, it is important to note that
Fig.~\ref{maps_means} depends on the observed energy $E_{\gamma,0}$, and more
specifically on the spectral shape of the factor $f_{\rm SUSY}$ in the energy
range $E_{\gamma,0}-E_{\gamma,0}(1+z_{\rm max})$, where $z_{\rm max}$ is the
largest redshift considered, in our case $z_{\rm max}=10$.  For most energies,
the $f_{\rm SUSY}$ spectrum is monotonically decreasing. In these regions, the
dependence on $I_{\gamma,0}$ with redshift will have the generic form shown in
Fig.~\ref{maps_means}. However, at the highest energies the shape is expected
to change dramatically due to the importance of Internal Bremsstrahlung and/or
of monochromatic lines (see Fig.~\ref{susy_energy}) close to the rest mass
energy of the dark matter particle. Such an effect is clearly visible in the
black dotted line in Fig.~\ref{maps_means}, which is the ReHS case for an
energy of $100\,{\rm GeV}$ and features a bump in the redshift distribution, a
reflection of the importance of IB at the highest energies.

The angular power spectrum of the EGB is an important tool to study the
statistical properties of its anisotropy on the sky. In fact, certain classes
of $\gamma$-ray sources are expected to exhibit different power spectra,
making this a potential means to identify the origin of the unresolved EGB.
If we decompose the relative deviations of the EGB around its mean into
spherical harmonics with coefficients $a_{lm}$,
\begin{equation}\label{harmonics}
  \Delta_{I_{\gamma,0}}(\theta,\phi)=\frac{I_{\gamma,0}(\theta,\phi)-\langle I_{\gamma,0}\rangle}{\langle
I_{\gamma,0}\rangle}=\sum_{l=0}^\infty\sum_{m=-l}^{m=l}a_{lm}Y_{lm}(\theta,\phi),
\end{equation}
then the angular power spectrum is defined by the real numbers
\begin{equation}\label{angular_power}
 C_l=\frac{1}{2l+1}\left(\sum_{m}\vert  a_{lm}\vert^2\right).
\end{equation}
For small angular scales (i.e.~large $l$), the variance of
$\Delta_{I_{\gamma,0}}$ per unit decade in $l$ can be connected directly to
the $C_l$'s through
\begin{equation}\label{connection}
\frac{\textrm{d}\langle\Delta^2_{I_{\gamma,0}}\rangle}{\textrm{dln}~l}\propto
l^2C_l  .
\end{equation}

\begin{figure}
\centering
\includegraphics[height=8.5cm,width=8.5cm]{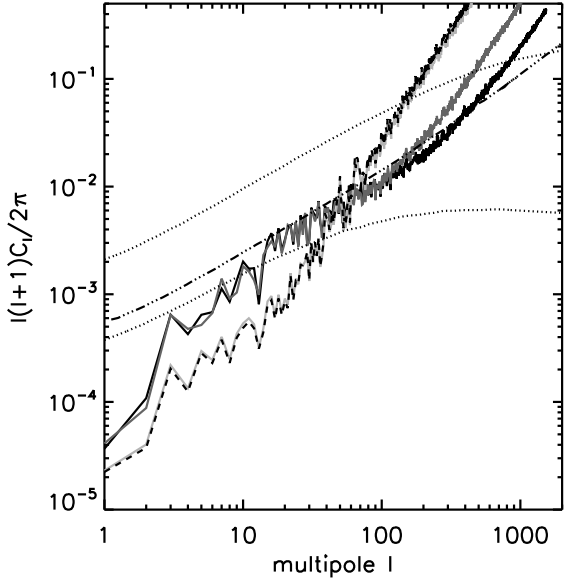}
\caption{Angular power spectrum of the EGB produced by dark matter
  annihilation as a function of the multipole $l$ for $E_{\gamma,0}=10\,{\rm
    GeV}$. The solid
  black line is for resolved haloes and subhaloes. The dashed black
  line is for main haloes only, resolved and unresolved down to the
  cutoff mass using a boost factor $b_h=60$. The light grey and dark grey
  solid lines show the contribution of all components, resolved and
  unresolved haloes and subhaloes, boosted with the extreme values in
  the interval $b_{\rm sub}=2-60$. The dotted lines show
  the predictions from \citet{Ando-07a} for a subhalo-dominant contribution
  with and without considering tidal destruction (lower and upper
  dotted lines), respectively. The dash-dotted line shows the case where
  the main halo dominates the signal according to the same study.}
\label{power}
\end{figure}

In Fig.~\ref{power}, we show the angular power spectrum $l(l+1)C_l/2\pi$ as
a function of the multipole $l$, at an observed energy of
$E_{\gamma,0}=10\,{\rm GeV}$, for all the cases discussed in
Fig.~\ref{maps_means}. At large scales, $l\lesssim10$, the power spectrum is
related to the clustering of dark matter haloes. All cases have the same shape
on these scales. However, when only the main halo contribution to the EGB is
considered (dashed black line), the normalization is lower because most of the
$\gamma$-ray signal comes from low mass haloes that are less clustered
(biased) than more massive haloes; recall that in this case we extrapolate the
signal down to the cutoff mass by assuming that the unresolved main
haloes are clustered in the same way as the least massive resolved haloes in
the MS-II.

The light grey line in Fig.~\ref{power}, corresponding to the full extrapolation
including subhaloes but with the minimal boost $b_{\rm sub}=2$, has exactly the
same power spectrum than the case with main haloes only, at all scales.  This
is because the signal from main haloes is dominant in this case, and subhaloes
have a negligible effect. In contrast, in the case where subhaloes have a
significant contribution mediated by $b_{\rm sub}=60$ (dark grey line), the
normalization is larger at small scales because the signal is dominated by
subhaloes belonging to the most massive haloes, and the latter are strongly
clustered.  

As the angular scale decreases, $l>10$, the power spectrum depends more and
more on the internal structures of haloes. For the cases where substructures
are ignored or are negligible (dashed-black and light grey lines), the slope becomes
steeper, with a slope close to 2. The power spectrum for the cases where
substructures are relevant (dark grey and black solid lines) behaves differently,
however.  In the range $l\in[20,100]$, it becomes slightly shallower, i.e.~the
signal is slightly more isotropic in this regime.  This is probably produced
by the distribution of substructure within the biggest haloes.  Contrary to
the strong central concentration of the matter in a halo, the number density
profile of subhaloes is considerably shallower than a NFW profile and produces
a luminosity profile in projection which is essentially flat
\citep{Springel-08b}.  This effect continues until $l\sim200$ where the power
spectrum becomes dominated by the low-mass main haloes.  The corresponding
upturn happens at smaller scales for the case corresponding to the black line
because the substructures in the most massive haloes remain important for a
larger angular range, whereas for the case of the dark grey line, the low-mass
unresolved main haloes start to dominate earlier.

The dash-dotted and dotted lines in Fig.~\ref{power} were taken from the
analysis of
\citet{Ando-07a}. They were obtained under the assumption of a Sheth-Tormen
mass function \citep{Sheth-Mo-Tormen-01}, a halo bias according to the model
of \citet{Mo-White-96}, and a NFW main halo density profile with a
concentration given by the mass-concentration relation found by
\citet{Bullock-01}.  Subhaloes were incorporated through a halo occupation
distribution model, where the number of subhaloes in each main halo, $\langle
N_{\rm sub}\vert M_{h}\rangle$, is assumed to depend only on the mass of the
host according to $\langle N_{\rm sub}\vert M_h\rangle\propto M_h^\beta$. The
upper dotted line in Fig.~\ref{power} is for $\beta=1.0$ and the lower dotted line for
$\beta=0.7$; in these cases subhaloes dominate completely the signal.
The subhaloes are assumed to be distributed with a NFW radial profile inside
the main halo. Since
they dominate the signal, the total gamma-ray luminosity (halo + subhaloes) is effectively
tracing the density of dark matter instead of the density squared as
expected for a smooth halo with no substructures. This case is represented
with a dash-dotted line in Fig.~\ref{power}. In all three cases only the
contribution of haloes down to a minimum mass of $10^6$M$_{\odot}$ is
taken into account.

There are several approximations used by \citet{Ando-07a} that we believe may
be the cause of
the differences between their results and ours. In our case, the clustering of
dark matter haloes is given directly by the simulations instead of by an
approximate halo model. This introduces differences that can only be
quantified by a direct comparison of the power spectrum of haloes used by
\citet{Ando-07a} and the one in the MS-II. Such a
comparison is out of the scope of this paper.
We need to mention however, that our simulated maps have a deficiency of
power at the largest angular scales due to the lack of
comoving scales larger than the simulation box (100 Mpc/h). This
is a problem already pointed out by  \citet{Carbone-08} that used a
similar map-making procedure, and is probably the main reason for the
disagreement between the results of \citet{Ando-07a} and ours on the large scale
regime. 

On the other hand, while we do assume a NFW profile, the concentration  of
each halo is given by its scaling properties (see Eqs.~\ref{delta_V} and
\ref{delta_c}), whereas \citet{Ando-07a} uses the mass-concentration fitting
relations found by \citet{Bullock-01}. This fitting formulae agree well with
simulations at $z=0$ but strongly disagree at higher redshift. They predict a
relatively strong correlation between concentration and mass, contrary to
simulations where this correlation gets flatter for higher redshifts. This
difference contributes to the discrepancy between our results and those of
\citet{Ando-07a} in the small scale regime. Finally, an important
disagreement is expected due to the treatment of subhaloes. The
model used by \citet{Ando-07a} produces a shallower power spectrum at small scales since the
gamma-ray luminosity of a halo is effectively given by the density of dark
matter instead of by the square of the density. In simulations, the radial distribution
of subhaloes is considerably shallower towards the centre than a NFW profile
due to tidal striping \citep[e.g.][]{Diemand-08,Springel-08b}. Near the halo
centre our expectation is that the signal
will be dominated by the smooth component rather than by substructures. This
is an effect that contributes to make our power
spectrum steeper at the smaller scales than the one obtained by \citet{Ando-07a}.

\subsection{Energy dependence}

\begin{figure}
\centering
\includegraphics[height=8.5cm,width=8.5cm]{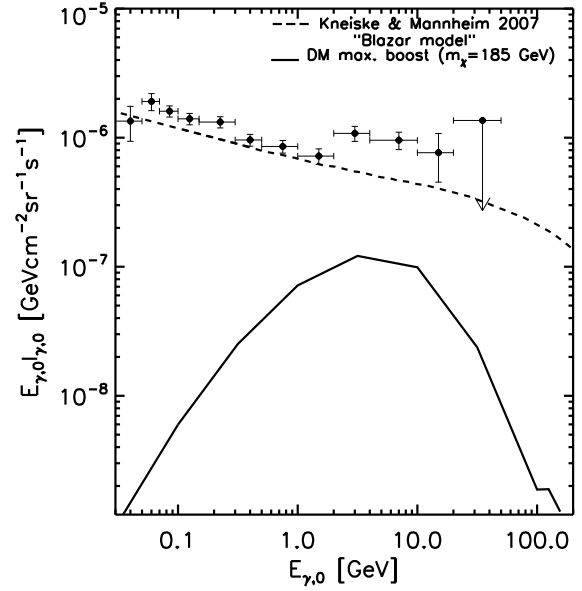}
\caption{Mean annihilation intensity $E_{\gamma,0}I_{\gamma,0}$ as a
  function of the observed energy for the maximum boost case,
  $b_h=60$, $b_{\rm sub}=60$.  The observational values and the error
  bars come from the EGRET satellite, as presented by
  \citet{Strong-Moskalenko-Reimer-04}.  The ``blazar model'' given by
  \citet{Kneiske-Mannheim-08} is shown as a dashed line.}
\label{energy_dep}
\end{figure}

So far we have restricted our analysis to an observed $\gamma$-ray energy of
$10\,{\rm GeV}$.  In the following we study how the isotropic and anisotropic
components of the $\gamma$-ray maps change with photon energy. The black solid
line in Fig.~\ref{energy_dep} shows the energy spectrum of the mean
annihilation intensity $I_{\gamma,0}$ for the case of the maximum boost
$b_h=60$, $b_{\rm sub}=60$ (corresponding to the dark grey line in
Fig.~\ref{maps_means}). For simplicity we refrain from showing the other cases
explicitly as they differ only in their normalization (for example, the
minimum boost case with $b_{\rm sub}=2$ simply lies 2 orders of magnitude
lower). The solid circles with error bars are observational measurements taken
by the satellite EGRET after removal of the galactic foregrounds
\citep{Strong-Moskalenko-Reimer-04}.  There are several sources other than DM
annihilation that contribute to the EGB in the range of energies shown in
Fig.~\ref{energy_dep}; blazars and cosmological structure formation shocks
are among the most important ones.  In Fig.~\ref{energy_dep}, we include the
contribution to the EGB from the blazar model introduced by
\citet{Kneiske-Mannheim-08} (dashed line) that explains most of the signal for
energies lower than $1\,{\rm GeV}$. However, there seems to be an excess of
$\gamma$-ray flux observed in the EGB for energies larger than $1\,{\rm GeV}$
that can not be readily explained by known astrophysical sources
\citep{Ong-06}. Interestingly, the $\gamma$-ray spectrum produced by dark
matter annihilation has the right shape to fit the observed values in this
energy range, even though this could of course be a coincidence.  Our maximum
boost scenario is in any case still too low (by about an order of magnitude) to
explain the dominant component of the EGB in this energy range, and an
additional boost factor would be needed to explain this signal, if confirmed,
with DM annihilation.  An improved and more precise measurement of the energy
spectrum will be made by FERMI very soon, so there is hope that this issue
will become clearer in the near future.  We note that completing and improving
the measurements for high energies $E_{\gamma,0}>20\,{\rm GeV}$ is of key
importance since the DM annihilation spectrum is expected to drop off sharply
at such higher energies, making this an important discriminant against other
types of sources.

The presence of a secondary bump at the highest energies,
$E_{\gamma,0}\sim125\,{\rm GeV}$ is a consequence of IB effects that become
important at these energies.  The relevance of this effect depends on the
specific parameters chosen for the mSUGRA model (see Fig.~\ref{susy_energy}),
but its detection in the isotropic component of the EGB is generically less
likely than identifying the primary bump.

\begin{figure}
\centering
\includegraphics[height=8.5cm,width=8.5cm]{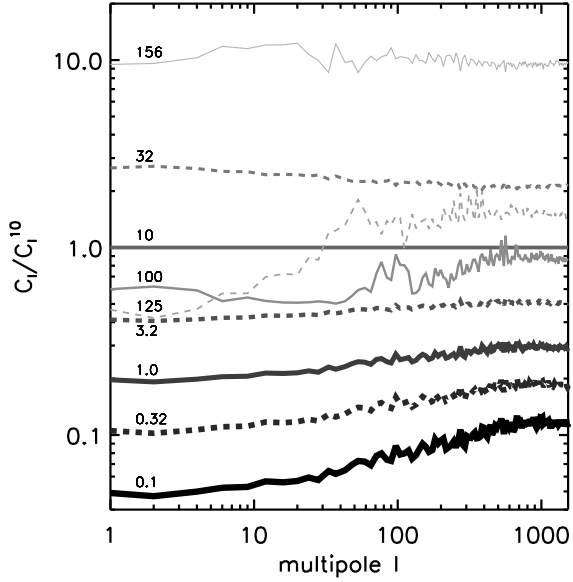}
\caption{Ratio of the angular power spectrum of the EGB produced by
  dark matter annihilation at different energies to the one at
  $E_{\gamma,0}=10\,{\rm GeV}$ (Fig.~\ref{power}). The lines are labeled
  according to their respective energy in GeV. They appear in a grey scale going
  from black to light grey, and with decreasing thickness, for increasing
  energy. The line styles alternate between solid and dashed for consecutive energies.}
\label{power_energy_dep}
\end{figure}

Fig.~\ref{power_energy_dep} shows the ratio of the power spectrum $C_l$ for
different energies to the power spectrum at $10\,{\rm GeV}$ as a function of
the multipole $l$.  The lines are labeled in the left of the figure according
to the energy of the map in GeV. For $E_{\gamma,0}\in(0.1,30)\,{\rm GeV}$ there is generally an
increase of power for increasing energy, a feature already pointed out by
\citet{Ando-Komatsu-06}.  Whereas overall the variations of the ratio
$C_l/C_{10}$ for these low-intermediate energies are small, there are some
differences at intermediate to small scales that have important consequences,
as we discuss below.  For the higher energies, $E_{\gamma,0}\in(100,156)\,{\rm
  GeV}$, the signal originates completely at relatively low redshifts. For
example at $E_{\gamma,0}=100\,{\rm GeV}$, {\em all} the signal comes from
$z<1.0$ (see Fig.~\ref{maps_means}) since the $f_{\rm SUSY}$ spectrum has a
sharp cut-off at the neutralino mass, $185\,{\rm GeV}$. The contribution of IB
to the photon spectrum is dominant at these energies.

We note that there is a prominent peak in the power spectra for
energies 100 and $125\,{\rm GeV}$, at $l=100$ and $l=60$,
respectively. This peak is in fact a spurious effect related to the
periodicity across the sky that is present within a given shell
(partial map). Since we are working with only one simulation (MS-II)
with a finite box size of $100\hMpc$, we are inevitably affected by
the periodicity of the replicas within a given shell.  For a
particular shell, this periodicity introduces a mild correlation at an
angular scale determined by the comoving distance from the observer to
the shell and the transverse box size of $100\hMpc$. This correlation
is largely averaged out for the full line of sight integration, as the
relevant angular scale changes with redshift.  However, for the
highest energies, the bump in the spectrum of $f_{\rm SUSY}$ maximizes
the contribution from a particular emitted energy, which in turn
translates into a particular redshift (see Fig.~\ref{maps_means}) that
contributes most prominently. As a result, the angular correlation
from the periodicity corresponding to this particular redshift is
particularly strong, and remains visible in the final power spectrum.

\begin{figure}
\centering
\includegraphics[height=8.5cm,width=8.5cm]{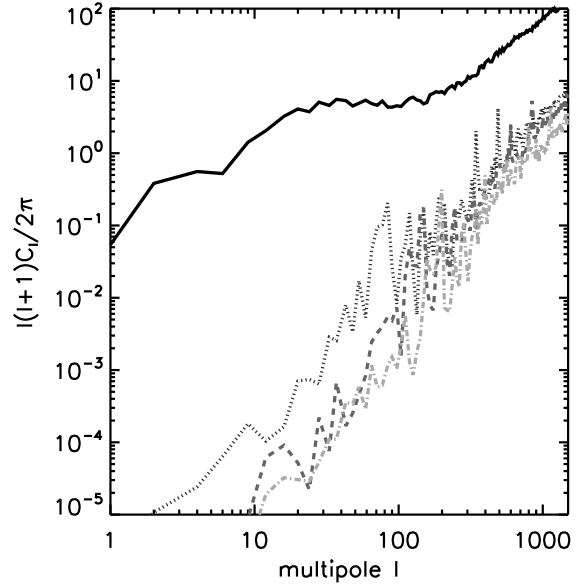}
\caption{Angular power spectrum of the EGB for different partial maps,
  i.e.~for the volume contained in a shell of comoving thickness
  $100\,h^{-1}{\rm Mpc}$ at a given redshift, for the sky map at
  observed energy $E_{\gamma,0}=10\,{\rm GeV}$.  The black solid, dark grey
  dotted, medium grey dashed
  and light grey dash-dotted lines are for $z=0$, $0.5$, $1.0$, and $1.5$ respectively.}
\label{partial_power}
\end{figure}

To make this effect more evident we show in Fig.~\ref{partial_power} the
angular power spectrum of different partial maps (i.e. single shells of
comoving thickness $100\,h^{-1}{\rm Mpc}$ along the line of sight), for an
observed energy of $E_{\gamma,0}=10\,{\rm GeV}$ (to show the periodicity
effect, the choice of energy is not important). The black solid, dark grey dotted,
medium grey dashed and light grey dash-dotted
lines are for shells at $z=0$, $0.5$, $1.0$, and $1.5$, respectively. Clearly,
there are peaks at certain values of $l$ for the red, green, and blue
lines. The values of the multipole $l$ where the peaks occur correspond to
angular scales that are closely related to the boxsize at the distance of the
shells.  Specifically we find 104.4, 116.7, 99.9 $\hMpc$ for $0.5$,
$1.0$, and $1.5$, respectively. This effect is not present at $z=0$, since the
first shell, with a depth of $50\hMpc$, encloses a volume too small to contain
the full box.

\begin{figure*}
\centering                      
\includegraphics[width=0.49\hsize]{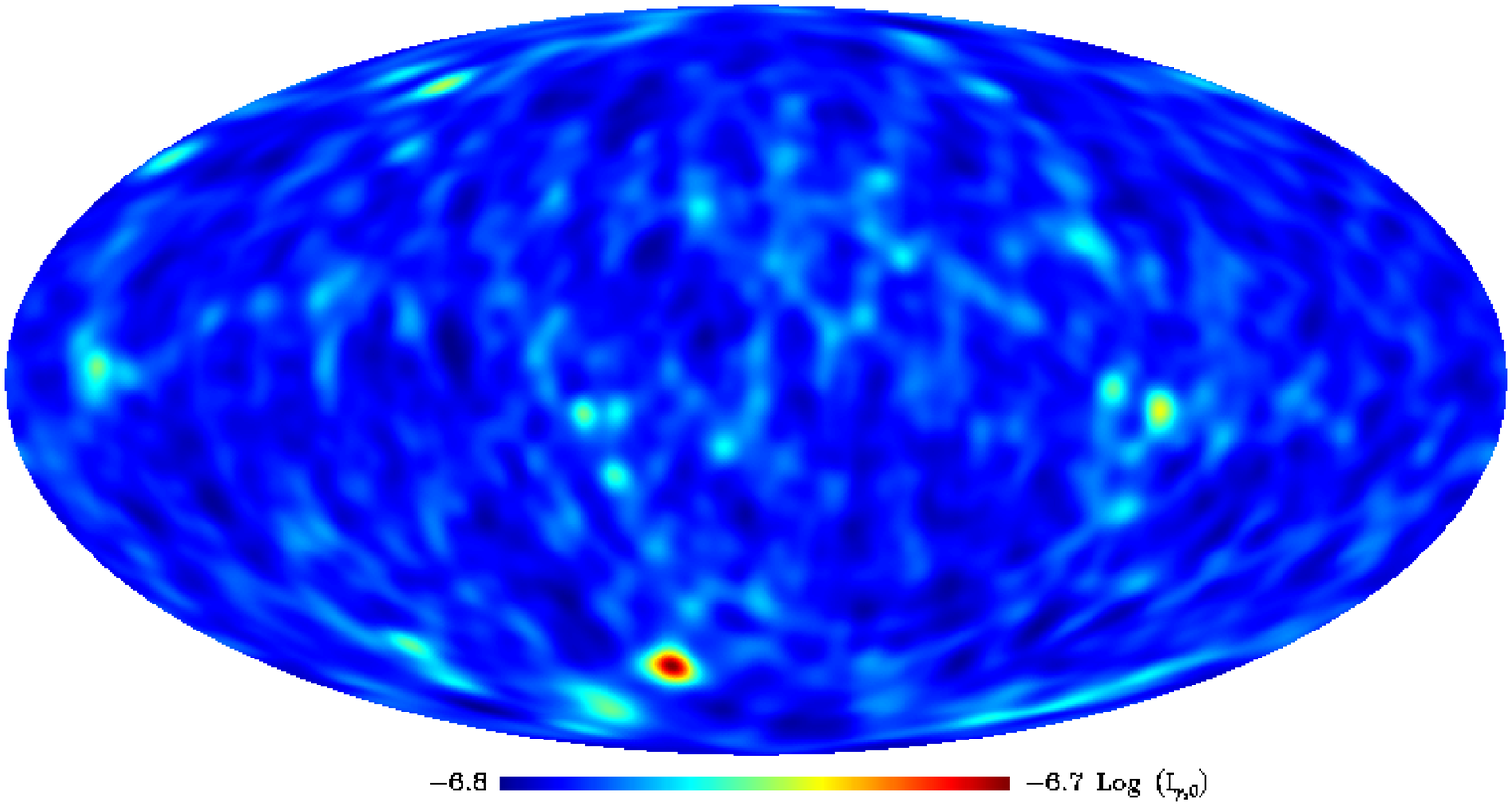}
\includegraphics[width=0.49\hsize]{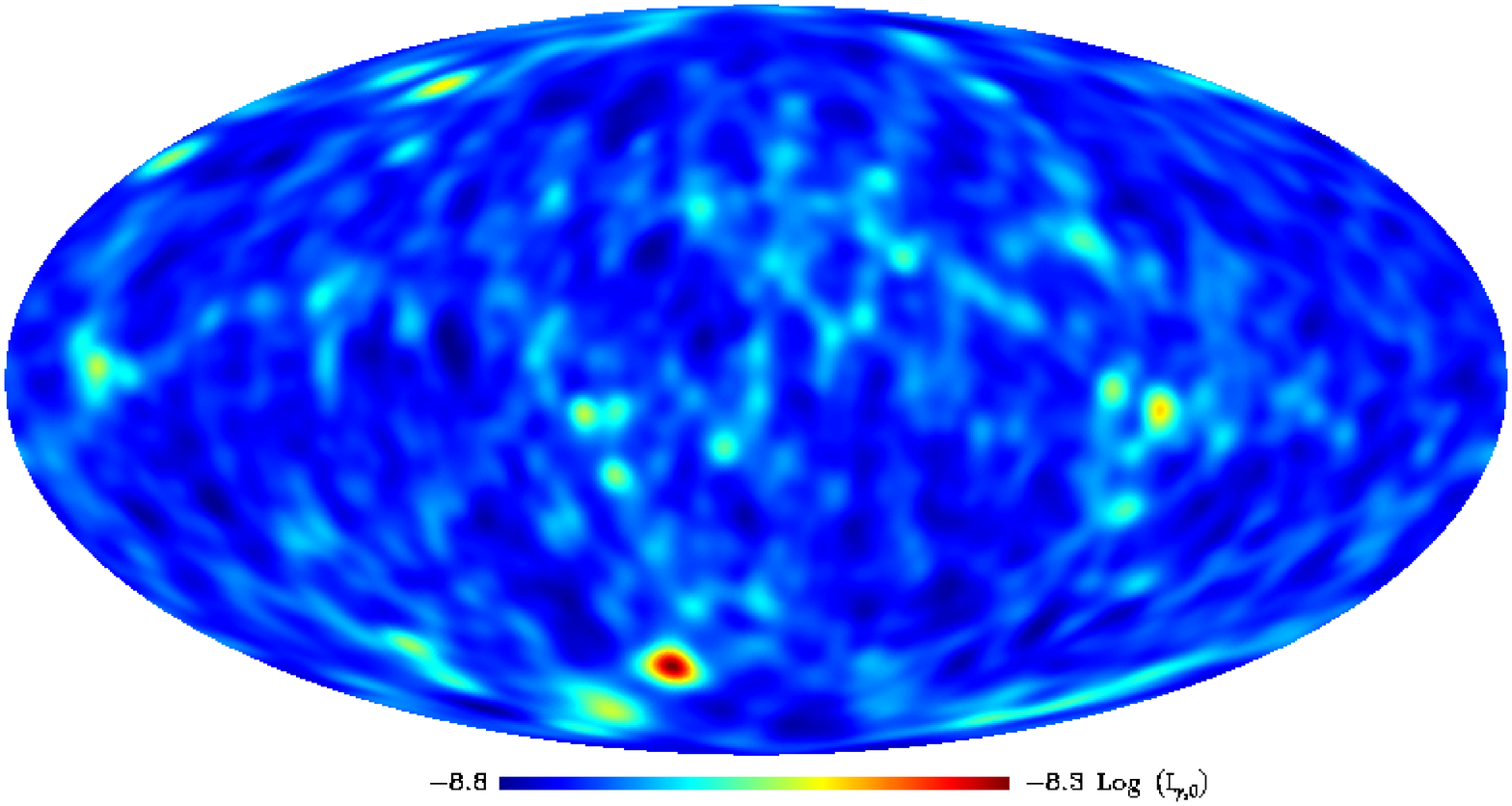}
\includegraphics[width=0.49\hsize]{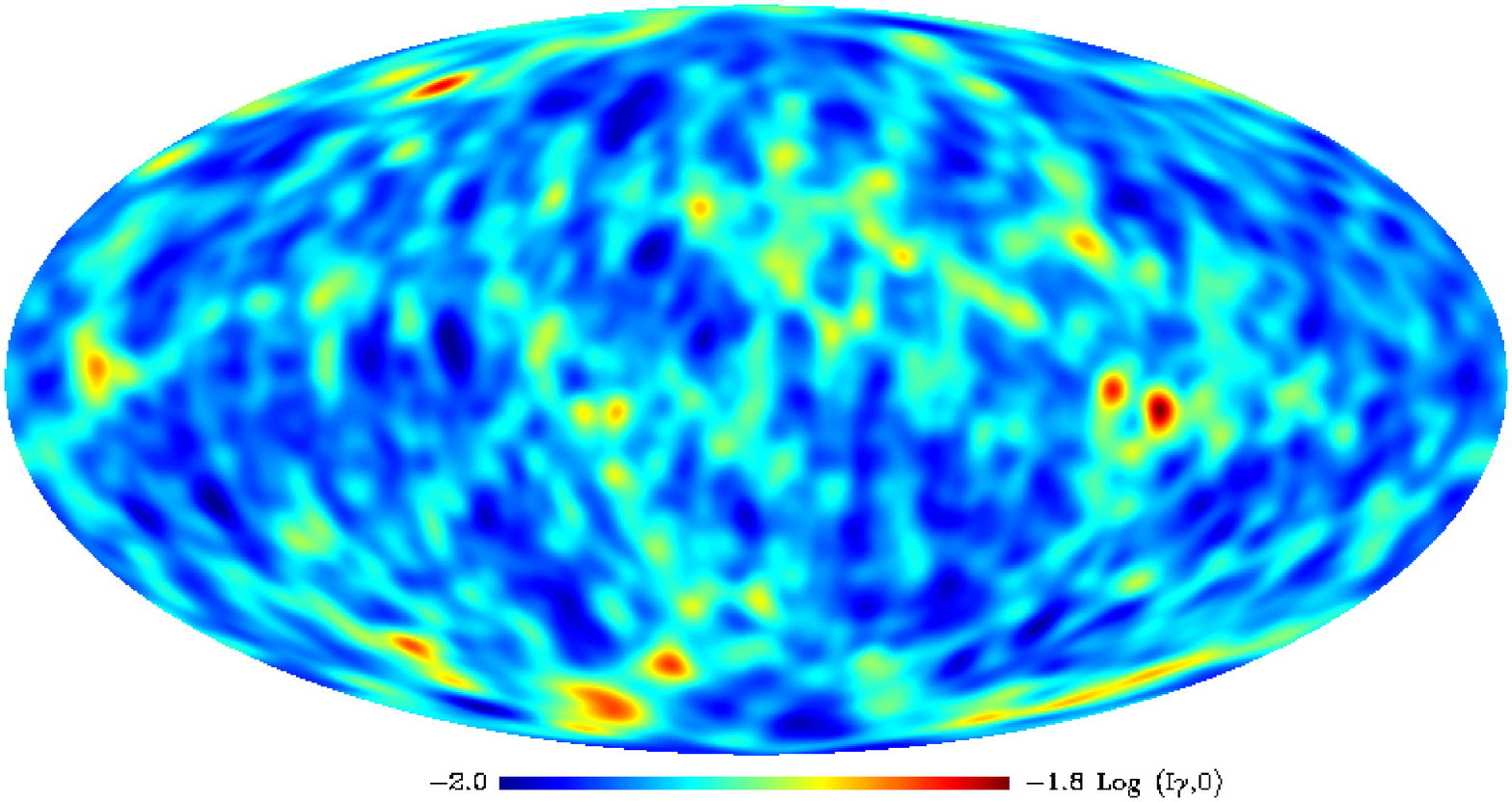}
\includegraphics[width=0.49\hsize]{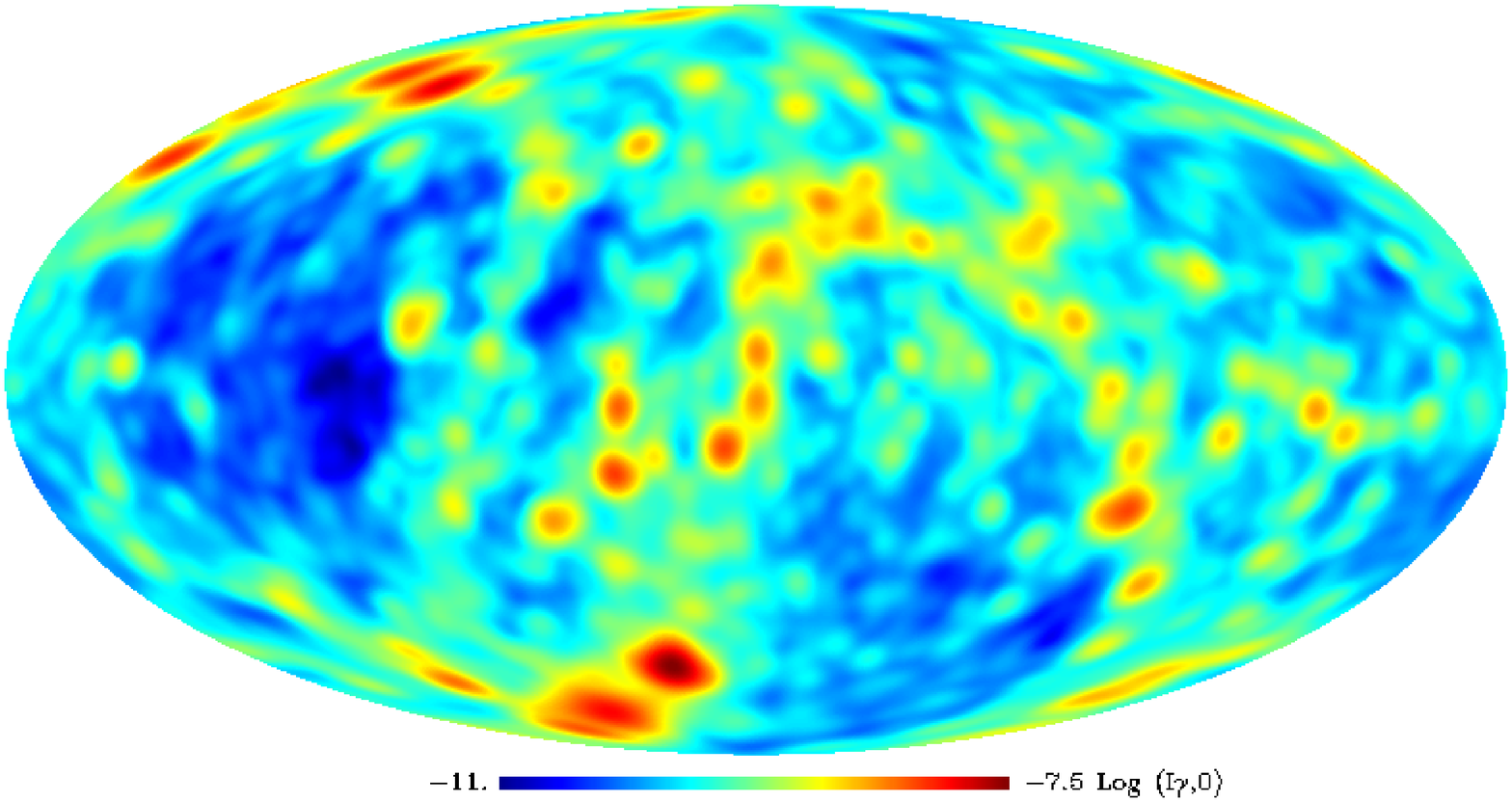}
\caption{{\it Upper panel:} Total sky maps at energies
  $E_{\gamma,0}=0.1\,{\rm GeV}$ and $E_{\gamma,0}=32\,{\rm GeV}$ in
  the left and right, respectively.  The maps were smoothed with a
  Gaussian beam with a FWHM of $5\degr$.  {\it Lower panel:} Ratio of
  the maps in the upper panel (left) and the partial map at $z=0$ for
  $E_{\gamma,0}=0.1\,{\rm GeV}$ (right).}
\label{smooth}
\end{figure*}

Another important result that we obtain from the analysis of the power spectra
is that there are differences in the shape of the power spectra in
Fig.~\ref{power_energy_dep} for different energies.  We can exploit these
differences and enhance the power at small or large angular scales by taking
the {\em ratio of sky maps at different energies}. In particular, the case of
the energies $E_{\gamma,0}=0.1\,{\rm GeV}$ and $E_{\gamma,0}=32\,{\rm GeV}$
(black-solid and red-dotted lines, respectively) is notable since these
energies are free from periodicity effects and when combined they can enhance
the intermediate to large angular scales. The upper panel of Fig.~\ref{smooth}
shows the total sky maps for $E_{\gamma,0}=0.1\,{\rm GeV}$ and
$E_{\gamma,0}=32\,{\rm GeV}$ in the left and right, respectively, smoothed
with a Gaussian beam with a FWHM of $5\degr$. Both maps show only small
anisotropies, with the exception of a couple of prominent structures. The sky
map in the lower left panel is the ratio of the two maps; it clearly enhances
the signal of nearby structures. This is even more convincing by looking at
the nearest partial map at $z=0$ for $E_{\gamma,0}=0.1\,{\rm GeV}$ shown in
the lower right corner of Fig.~\ref{smooth}.  Most of the prominent dark
matter structures that can be seen in this map are also clearly present in the
map with the energy ratio.  We conclude that the $\gamma$-ray sky maps that
the FERMI satellite will obtain at different energies could be used to construct difference
maps which may then show enhanced correlations with nearby cosmic large-scale
structure. 

In our results, we have neglected the interactions between the
gamma-ray photons produced by annihilation and the lower energy
photons from intervening starlight produced in galaxies. This
interaction enters effectively as an absorption term in the specific
intensity in Eq.~(\ref{intensity}) due to pair production, and is
usually parameterised by an exponential factor
{\small $e^{-\tau(z,E_{\gamma,0})}$}
\citep[e.g.][]{Ando-Komatsu-06,Fornasa-09}. \citet{Fornasa-09} give
an approximate expression for the opacity $\tau$ following the work
of \citet{Salamon-Stecker-98}:
{\small $\tau(z,E_{\gamma,0})=z/3.3(E_{\gamma,0}/10{\rm
    GeV})^{-0.8}$}. The absorption term is typically small for
energies $\lesssim100$GeV. We carried out the exercise of producing
a pair of sky maps for energies of 10 and 100 GeV including the
absorption term following the parameterisation of
\citet{Fornasa-09}. For $E_{\gamma,0}=10$GeV we found no significant
changes compared to the map without absorption; the total value of
$I_{\gamma,0}$ drops by less than $10\%$ and the normalization of
the angular power spectrum increases by $\sim20\%$ preserving the
same shape. For $E_{\gamma,0}=100$GeV the differences are more
relevant, the mean specific intensity decreases by $50\%$, the bump
shown in the top panel of Fig.~\ref{maps_means} (dotted line) becomes
less pronounced and therefore the angular power spectrum gets closer
in shape to the ones at lower energies; this is an expected effect
since absorption is stronger at higher redshifts and thus reduces
the contribution of the ``energy bump" at this particular map.

\subsection{Additional boost factor} 

The role and importance of annihilations of neutralinos for the EGB is
subject to many uncertainties, related to both the intrinsic nature of
neutralinos (as a supersymmetric particle) and to its spatial
distribution (as dark matter) in the different structures in the
Universe.  The former uncertainties are ultimately connected to the
assumed supersymmetric model, and within the mSUGRA paradigm, are
approximately constrained by the results shown in
Fig.~\ref{susy_energy}. In our calculations, they are encapsulated in
the value of the quantity $f_{\rm SUSY}$. We chose a particular point,
`L', in the allowed mSUGRA parameter space as a fiducial case to
obtain quantitative results.  This benchmark point is close to be the
upper bound for the value of $f_{\rm SUSY}$ and hence represents a
best case scenario for a ``high'' value of the annihilation signal.
This is important in connection with the results obtained for the
expected specific intensity spectrum (Fig.~\ref{energy_dep}), where
the value for each energy is given by the isotropic component of our
simulated sky maps.  We mentioned earlier that in order to match the
observational results coming from the EGRET satellite in the
$1-10\,{\rm GeV}$ region, a further ``boost'' factor of $\sim10$
appears necessary. Since the SUSY benchmark point L is close to an
upper limit to the value of $f_{\rm SUSY}$, this boost factor can
probably not be accounted for by a different choice of parameters in
the mSUGRA parameter space.  Of course, the situation could be
different in broader classes of SUSY theories outside the mSUGRA
paradigm, but is certainly true for our analysis.

We note that there are a number effects we have not accounted for but
which could possibly give such an additional boost factor: (i) We have
not incorporated the effects of forming luminous galaxies onto the
dark haloes; it is generally believed that the formation of a baryonic
disk contracts the halo, pulling matter inwards towards the centre
\citep[e.g.][]{Mo-Mao-White-98,Gnedin-04}, yielding a higher
concentration of the dark matter and a boost of the annihilation flux;
(ii) supermassive black holes could produce ``spikes'' of dark
matter in the galactic centre
\citep{Gondolo-Silk-99,Bertone-Merritt-05a}. Such spikes may also
yield an enhanced annihilation signal.  (iii) mini-spikes of dark
matter are expected to form around intermediate-mass black holes
(IMBHs) (with masses in the range $10^2\lesssim M/M_{\odot}\lesssim
10^6$); in certain scenarios, the gamma-ray luminosity of each
mini-spike could be as large as the luminosity of the entire halo. For
a typical Milky Way halo, the number of IMBHs is expected to be $\sim
100$ \citep{Bertone-Zentner-Silk-05}; (iv) the non linear
collapse of dark haloes for cold collisionless dark matter leads to
the formation of caustics
\citep[e.g.][]{Hogan-01,White-Vogelsberger-09}, which, due to their
high density can lead to a large annihilation rate and gamma-ray
luminosity. However, \citet{Vogelsberger-09a}, using simulations of
isolated haloes under self-similar and spherically symmetric initial
conditions, have recently found that the contribution of caustics to
the annihilation flux is less than previously expected. Based on
their results, caustics are expected to produce at most a global
annihilation flux comparable to the one from the smooth halo
component; (v) finally, the so-called Sommerfeld enhancement, a
quantum mechanical focusing effect, may lead to an effective increase
of the cross section for dark matter with low velocity dispersion,
such an enhancement could lead to a boost factor of several orders of
magnitude \citep{Lattanzi-Silk-09,Bovy-09}.
 
\section{Conclusions}

Neutralinos in supersymmetric extensions of the standard model of particle
physics are one of the most promising candidates for being the dominant
component of dark matter in the Universe. One attractive method for their
detection is to look for the products of their mutual annihilations. Among
such products are $\gamma$-rays that could be produced abundantly enough in
the dense centres of dark matter haloes to be susceptible of detection.  In
this work, we have studied the contribution of these $\gamma$-rays to the EGB
radiation. Excluding the contributions from the Milky Way halo itself, all
dark matter haloes and their substructures within the past light cone of an
observer in our galaxy contribute to the EGB.  The main goal of this paper has
been to calculate the expected emission from this light-cone on the basis of
the state-of-the-art Millennium-II simulation of cosmic structure formation
\citep{Boylan-Kolchin-09}.

The $\gamma$-ray luminosity of each halo (or subhalo) in the simulation can be
conveniently separated into two factors: (i) the SUSY factor $f_{\rm SUSY}$,
encompassing the intrinsic properties of neutralinos as SUSY particles (like
its mass, cross section and photon yield), and (ii) the astrophysical factor
related to the density squared dependence of the annihilation rate and
therefore to the clustering of neutralinos as dark matter.

We have discussed in some detail the first of these factors in Section 2, with
the aim of highlighting the energy dependence of $f_{\rm SUSY}$ for a
particular set of benchmark points within the mSUGRA framework.  We also
discussed the general mechanisms of photon production: decay of neutral pions
produced during the hadronization of the primary annihilation products; (ii)
loop-suppressed two-body final states containing photons; and (iii) internal
bremsstrahlung (IB).  Among the different benchmark points analyzed
we found an uncertainty of three orders of magnitude in the normalization of
$f_{\rm SUSY}$. We chose, for definiteness, one of the benchmark points with
a high photon yield and a neutralino mass of $m_{\chi}=185 \,{\rm GeV}$ for our
subsequent analysis.

We then analyzed the astrophysical factor in terms of the signals expected
from main haloes and their embedded subhaloes. In order to avoid strong
resolution effects, the internal structure of haloes was approximated by
spherically symmetric NFW density profiles, allowing an analytic integration
of the total luminosity expected from each halo. This effectively yields a
scaling law where the flux from a halo can be determined by two of its
structural properties alone, $V_{\rm max}$ and $r_{\rm max}$.  Using this
scaling law together with the abundance of haloes per unit mass range as given
by the MS-II simulation, we obtained a power law description for the
contribution of main haloes of a given mass interval to the total $\gamma$-ray
luminosity in a cosmic volume. An extrapolation of this power law to the
damping scale limit of neutralinos in the $\sim100\,{\rm GeV}$ mass range
allowed us to include the contribution of unresolved main haloes to the total
gamma-ray luminosity.  A similar universal power law was established for
describing the contribution of subhaloes to the luminosity of their host by
analyzing the most massive haloes in the MS-II containing at least 500
subhaloes. Using reasonably conservative error bounds for the large
extrapolation undertaken here, we found that the ``boost'' factor due to
substructures lies somewhere in the range 2 to $\sim 1800$ for a MW-size halo.
With these results in mind, we note that the uncertainty in the absolute
normalization of the astrophysical factor in the annihilation contribution to
the EGB amounts to two to three orders of magnitude.

We implemented a new map-making procedure that for the first time allowed us
to construct realistic realizations of the expected extragalactic $\gamma$-ray
sky from dark matter annihilation.  The angular resolution of the maps was
chosen to be close to the angular resolution of the recently launched FERMI
satellite, $\sim 0.115\degr$. We found that structures up to $z\sim 2$
contribute significantly in the low to intermediate range of the energy
spectrum.  In particular, for an observed $\gamma$-ray energy of $10\,{\rm
  GeV}$, dark matter structures beyond $z\sim 2$ contribute less than $5\%$ to
the total flux. At the highest energies, between $100-185\,{\rm GeV}$ for the
particular benchmark point we chose, the IB contribution to the photon yield
is dominant, resulting in EGB contributions that are predominantly coming from
a narrow redshift range, followed by a sharp cut-off towards higher redshifts
that reflects the cut-off in the energy spectrum of the $f_{\rm SUSY}$ factor.

Adopting maximum values for the expected signal boost due to substructures, we
found that our prediction for the energy spectrum of the isotropic component
of the EGB from dark matter annihilations lies approximately one order of
magnitude below the observed values of the EGB according to the satellite
EGRET, after foreground removal, in the energy range $1-20\,{\rm GeV}$, where
an apparent excess of $\gamma$-rays has been reported.  The latter has been
interpreted by several authors in the past as a possible signal of dark matter
annihilation.  At the highest energies, IB produces a small bump in the energy
spectrum that could in principle be observable but that lies orders of
magnitude lower than the expected background radiation for the particular
benchmark point analyzed.

We studied the anisotropic component of the EGB by computing the angular
power spectrum of the simulated maps. This yielded specific predictions for
the shape of the power spectrum, which can potentially be used to discriminate
against other sources of $\gamma$-rays, since the annihilation signal depends
in a specific and unique way on the large scale distribution of haloes, on the
distribution of subhaloes within haloes, and on the abundance and internal
structure of haloes as a function of time. The shape of the power spectrum was
found to depend on the energy of the observations. Interestingly, these
differences can be exploited to construct `color' maps that enhance the signal
of nearby dark matter structures, akin to hardness ratio maps in X-ray
observations. For example, we found that taking the ratio of the maps at
energies of $0.1\,{\rm GeV}$ and $32\,{\rm GeV}$ greatly enhances the contrast
of local dark matter structures, making them clearly visible in the
$\gamma$-ray sky. If strong spectral features in the rest-frame emission
spectrum of the annihilation radiation are present, this could be especially
powerful, perhaps even allowing tomographic observations of dark matter
structures.

Although the subtraction of foreground radiation and other astrophysical
sources contributing to the EGB is a difficult task, the specific shape of the
contribution of dark matter annihilation to the power spectrum of the EGB
makes this statistical tool attractive for potentially detecting dark matter
in the near future \citep[as has been pointed out previously by,
e.g.,][]{Ando-Komatsu-06}. The maps we constructed in this work are a
particular useful tool to help designing and testing data analysis methods for
uncovering this signal from real data. Also, we have here been able to show
that the energy dependence of the signal offers an additional highly
attractive way to detect specific signatures of the annihilation emission
spectrum. It is quite possible that the different energy channels onboard the
satellite FERMI that is currently mapping the EGB will already allow a
fruitful application of such an approach.

\section*{Acknowledgments}

JZ would like to thank Simon D. M. White, Mark Vogelsberger and Dar\'{i}o 
N\'u\~nez for interesting and critical
discussions. JZ is supported by the Joint Postdoctoral Program in Astrophysical Cosmology of the
Max  Planck Institute for Astrophysics and the Shanghai Astronomical
Observatory. Some of the results in this paper have been derived using
the HEALPiX \citep{Gorski-05} package.

\bibliography{./lit}

\label{lastpage}

\end{document}